\documentclass[]{rsif}

\jname{rsif}
\Journal{J. R. Soc. Interface }

\usepackage[usenames,dvipsnames]{xcolor}
\usepackage{graphicx,multirow,dcolumn,hyperref}
\usepackage{adjustbox,xspace,datetime,lipsum,tikz,soul}
\usepackage[textsize=scriptsize,textwidth=1.75cm,shadow]{todonotes}

\DeclareGraphicsExtensions{.pdf,.eps,.png,.jpg}

\definecolor{blue}{RGB}{0,85,142}
\definecolor{green}{RGB}{77,143,42}
\definecolor{plum}{RGB}{106,40,134}
\definecolor{gray}{RGB}{95,95,95}

\newcommand{\cmp}[1]{\mathcal{O}(#1)}

\newcommand{\conv}[1]{\mathcal{H}(#1)}
\newcommand{\neigh}[1]{\Gamma_{#1}}
\newcommand{\avg}[1]{\langle#1\rangle}

\newcommand{\set}[1]{\{#1\}}

\DeclareRobustCommand{\diams}[0]{%
	\tikz{\fill[green] (0.707ex,0) -- (1.414ex,0.707ex) -- (0.707ex,1.414ex) -- (0,0.707ex) -- cycle;
	\draw[thick] (0.707ex,0) -- (1.414ex,0.707ex) -- (0.707ex,1.414ex) -- (0,0.707ex) -- cycle;}\xspace}
\DeclareRobustCommand{\diame}[0]{%
	\tikz{\fill[white] (0.707ex,0) -- (1.414ex,0.707ex) -- (0.707ex,1.414ex) -- (0,0.707ex) -- cycle;
	\draw[thick,green] (0.707ex,0) -- (1.414ex,0.707ex) -- (0.707ex,1.414ex) -- (0,0.707ex) -- cycle;}\xspace}
\DeclareRobustCommand{\squrs}[0]{%
	\tikz{\fill[plum] (0,0) rectangle (1ex,1ex);\draw[thick] (0,0) rectangle (1ex,1ex);}\xspace}
\DeclareRobustCommand{\squre}[0]{%
	\tikz{\fill[white] (0,0) rectangle (1ex,1ex);\draw[thick,plum] (0,0) rectangle (1ex,1ex);}\xspace}
\DeclareRobustCommand{\trians}[0]{%
	\tikz{\fill[blue] (0,0) -- (1.25ex,0) -- (0.625ex,1.25ex) -- cycle;%
	\draw[thick] (0,0) -- (1.25ex,0) -- (0.625ex,1.25ex) -- cycle;}\xspace}
\DeclareRobustCommand{\triane}[0]{%
	\tikz{\fill[white] (0,0) -- (1.25ex,0) -- (0.625ex,1.25ex) -- cycle;%
	\draw[thick,blue] (0,0) -- (1.25ex,0) -- (0.625ex,1.25ex) -- cycle;}\xspace}
\DeclareRobustCommand{\ellps}[0]{%
	\tikz{\fill[gray] (0,0) circle (0.5ex);\draw[thick] (0,0) circle (0.5ex);}\xspace}
\DeclareRobustCommand{\ellpe}[0]{%
	\tikz{\fill[white] (0,0) circle (0.5ex);\draw[thick,gray] (0,0) circle (0.5ex);}\xspace}

\newcommand{\cref}[1]{reference~\cite{#1}\xspace}
\newcommand{\crefs}[1]{references~\cite{#1}\xspace}
\newcommand{\secref}[1]{section~\ref{sec:#1}\xspace}
\newcommand{\Secref}[1]{Section~\ref{sec:#1}\xspace}
\newcommand{\figref}[1]{figure~\ref{fig:#1}\xspace}
\newcommand{\Figref}[1]{Figure~\ref{fig:#1}\xspace}

\newcommand{\tblref}[1]{table~\ref{tbl:#1}\xspace}
\newcommand{\Tblref}[1]{Table~\ref{tbl:#1}\xspace}

\renewcommand{\eqref}[1]{equation~(\ref{eq:#1})\xspace}

\newcommand{\appref}[1]{appendix~\ref{app:#1}\xspace}
\newcommand{\mc}[1]{\multicolumn{1}{c}{#1}}

\newcommand{\jazz}{Jazz musicians\xspace}
\newcommand{\netsci}{Network scientists\xspace}
\newcommand{\cmpsci}{Computer scientists\xspace}
\newcommand{\celeg}{\emph{Caenorhabditis elegans}\xspace}
\newcommand{\scere}{\emph{Saccharomyces cerevisiae}\xspace}
\newcommand{\plasm}{\emph{Plasmodium falciparum}\xspace}
\newcommand{\oreg}[1]{AS (January 1, #1)\xspace}
\newcommand{\littlerock}{Little Rock Lake\xspace}
\newcommand{\baywet}{Florida Bay (wet)\xspace}
\newcommand{\baydry}{Florida Bay (dry)\xspace}
\newcommand{\ncsst}{\mc{N} & \mc{CS} & \mc{ST}}
\newcommand{\csst}{\mc{CS} & \mc{ST}}

\begin{document}

%
%

\title{Convex skeletons of complex networks}

\author{Lovro \v{S}ubelj}
\address{University of Ljubljana, Faculty of Computer and Information Science, Ljubljana, Slovenia}
\corres{Lovro \v{S}ubelj\\\email{lovro.subelj@fri.uni-lj.si}}

\subject{mathematical physics}

%
%

\begin{abstract}
	A convex network can be defined as a network such that every connected induced subgraph includes all the shortest paths between its nodes. Fully convex network would therefore be a collection of cliques stitched together in a tree. In this paper, we study the largest high-convexity part of empirical networks obtained by removing the least number of edges, which we call a convex skeleton. A convex skeleton is a generalisation of a network spanning tree in which each edge can be replaced by a clique of arbitrary size. We present different approaches for extracting convex skeletons and apply them to social collaboration and protein interactions networks, autonomous systems graphs and food webs. We show that the extracted convex skeletons retain the degree distribution, clustering, connectivity, distances, node position and also community structure, while making the shortest paths between the nodes largely unique. Moreover, in the Slovenian computer scientists coauthorship network, a convex skeleton retains the strongest ties between the authors, differently from a spanning tree or high-betweenness backbone and high-salience skeleton. A convex skeleton thus represents a simple definition of a network backbone with applications in coauthorship and other social collaboration networks.
\end{abstract}

%
%

\begin{fmtext} \section{\label{sec:intro}Introduction}

An integral part of modern network analysis is understanding the structural properties of real empirical networks through different statistical models and techniques. One such example is the study of shortest paths between the nodes in a network, which has proven very useful in the past. Milgram's small-world experiment is a well-known empirical example~\cite{Mil67}. As another example, the frequently used measure of betweenness centrality is also based on the number of shortest paths~\cite{Fre77}.

A related, but different concept is that of convexity studied in metric graph theory~\cite{HN81,FJ86,Van93,Pel13}. Convexity is a property of a part of a mathematical object that includes

\end{fmtext} \maketitle

\noindent all the shortest paths between its units. In the case of graphs or networks, a connected induced subgraph is said to be convex if every shortest path between the nodes of the subgraph lies entirely within the subgraph. Note that while convexity demands the inclusion of all shortest paths between the nodes of the subgraph it makes no implications regarding their length nor number.

Two methodologies for analysing convexity in networks have been proposed in the literature. Everett and Seidman~\cite{ES85} define the hull number of a network as the size of the smallest subset of nodes whose convex hull is the entire network. Here, convex hull of a subset of nodes is the smallest convex subgraph including those nodes. Furthermore, Marc and \v{S}ubelj~\cite{MS18} define network convexity through the expansion of randomly grown subsets of nodes to convex subgraphs. In contrast to the hull number, which is an NP-hard problem~\cite{DGKPS09}, network convexity has a polynomial computational complexity and has already been successfully used in characterising the structure of empirical networks. In particular, it measures whether a network is either tree-like or clique-like on all different scales.

In this paper, we study the largest high-convexity part of empirical networks obtained by removing the least number of edges, which we call a convex skeleton. As demonstrated below, a convex skeleton is a collection of cliques or nearly cliques that are stitched together in a tree-like arrangement. We present different approaches for extracting convex skeletons and apply them to social collaboration and protein interactions networks, autonomous systems graphs and food webs. We show that the extracted convex skeletons retain most important structural properties of networks, differently from some popular network backboning techniques~\cite{HLMSW16,CN17} such as a spanning tree or high-betweenness backbone~\cite{Fre77} and high-salience skeleton~\cite{GTB12}. At the same time, a convex skeleton implies a very simple structure consisting only of cliques and a tree. It seems to provide a particularly reasonable abstraction of social collaboration networks, with applications in coauthorship networks and also elsewhere.

The rest of the paper is structured as follows. In~\secref{convex}, we provide a formal definition of network convexity and review the main results in~\cref{MS18} relevant for this work. Next, in~\secref{random}, we study the robustness of convexity under random perturbations of network structure. \Secref{skeleton} introduces the concept of a convex skeleton and presents different approaches for extracting convex skeletons from empirical networks. These are applied to networks of various type and origin in~\secref{skeleton}, while in~\secref{apps} we more thoroughly analyse node position and community structure of convex skeletons, and compare different backbones extracted from the Slovenian computer scientists coauthorship network. \Secref{conc} concludes the paper and gives some final remarks on convex skeletons.

%
%

\section{\label{sec:convex}Convexity in networks}

The definition of convexity in networks adopted here is based on the concept of a convex subgraph~\cite{HN81,FJ86,Van93,Pel13}. Let $S$ be a connected subset of nodes in a network. The subgraph induced by the nodes in $S$ is convex if all the shortest paths or geodesics between any two nodes in $S$ are entirely included within $S$. For instance, in a tree or a complete graph, every connected subset of nodes $S$ induces either a convex subtree or a clique, which is always convex. Note that convex subgraphs must include all the shortest paths between its nodes, whereas a subgraph including at least one shortest path between each pair of nodes is called an isometric subgraph. 

\begin{figure}[t]
	\centering\includegraphics[width=0.6\textwidth]{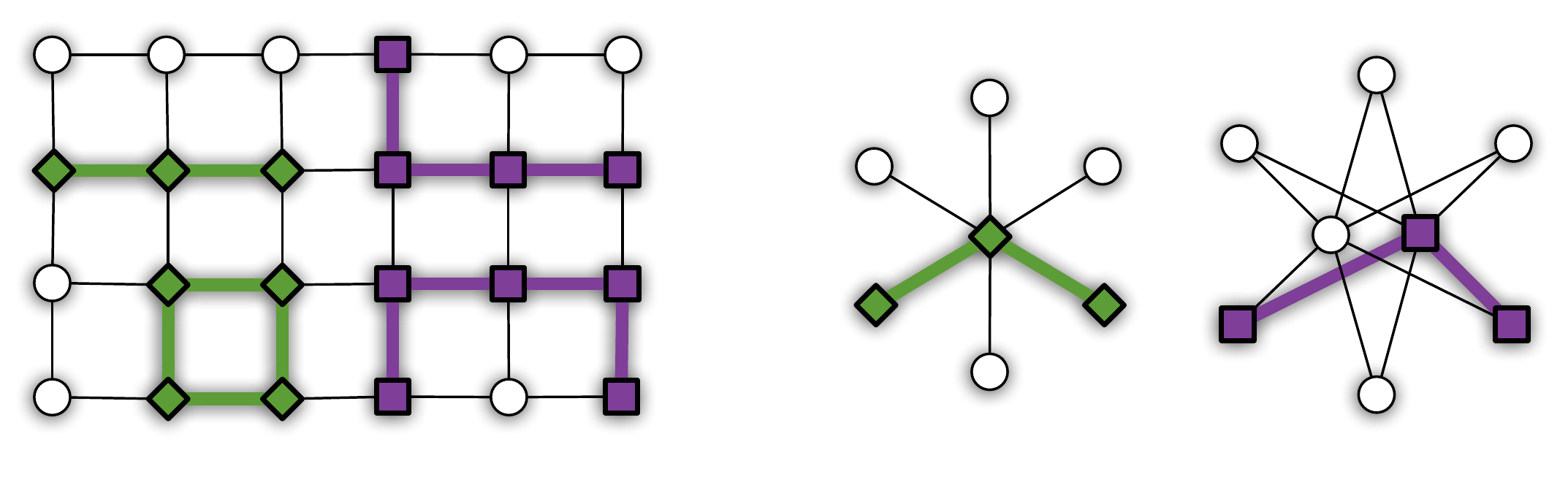}%
	\caption{\label{fig:examples}(\emph{left})~Rectangular lattice with highlighted convex~(\diams) and non-convex (\squrs) subgraphs. (\emph{right})~Two star graphs in which every connected triplet of nodes induces either a convex or non-convex subgraph, respectively.}
\end{figure}

The left side of~\figref{examples} shows different examples of convex and non-convex subgraphs of a rectangular lattice, where
any convex subgraph is necessarily a rectangular sublattice. Furthermore, the right side of~\figref{examples} shows two seemingly similar, but different graphs consisting of a star graph and a two-star arrangement. While any connected induced subgraph is convex in the star graph, no non-trivial subgraph is convex in the second graph. Convex subgraphs thus characterise networks beyond connected and induced subgraphs~\cite{Bat88}, also known as motifs~\cite{MSIKCA02} and graphlets~\cite{PCJ04} in networks jargon.

To ease the terminology, we say that a connected subset of nodes $S$ is convex if it induces a convex subgraph in a given network. Now, assume the subset $S$ is not convex. The smallest convex subgraph including all the nodes in $S$ is called the convex hull $\conv{S}$~\cite{HN81,FJ86}, which is uniquely defined. Obviously, $\conv{S}=S$ only when $S$ is a convex subset. For example, the convex hull $\conv{S}$ of any connected triplet of nodes $S$ in the two-star graph in~\figref{examples} spans the whole graph, whereas $\conv{S}=S$ for any connected subset $S$ in the star graph.

The above suggests a possible definition of convexity in networks~\cite{MS18}. One randomly grows connected induced subgraphs or subsets of nodes $S$ one node at a time and expands them to their convex hulls $\conv{S}$ if needed. Any subset $S$ generated by this expansion procedure therefore induces a convex subgraph. By observing the growth of subsets $S$, one can measure convexity in a network. In networks in which any connected induced subgraph is likely to be convex, the subsets $S$ would grow slowly, a single node at a time. Such networks are termed convex networks~\cite{MS18}. On the other hand, in non-convex networks such as food webs and random graphs as we show below, the subsets $S$ expand rapidly and include all the nodes in a network after only a few steps of the procedure.

The details of the convex expansion procedure are as follows. One starts by initialising the subset $S$ with a randomly selected seed node. Then, on each step of the procedure, a new node is chosen by following a random edge leading outside of $S$. Hence, a node $i\notin S$ is selected with the probability $\propto|\neigh{i}\cap S|$, where $\neigh{i}$ is the set of its neighbours. Last, the newly selected node $i$ is added to the subset $S$, which is finally expanded to a convex subset as $\conv{S\cup\set{i}}$. The entire process is repeated until the subset $S$ includes all the nodes in a network. We highlight that neither the selection of the initial seed node nor the subsequent choices of including new nodes affect the evolution of the procedure~\cite{MS18}. The algorithmic complexity of a single run in a network with $n$ nodes and $m$ edges is $\cmp{nm}$~\cite{convex}.

\begin{figure}[b]
	\centering\includegraphics[width=0.75\textwidth]{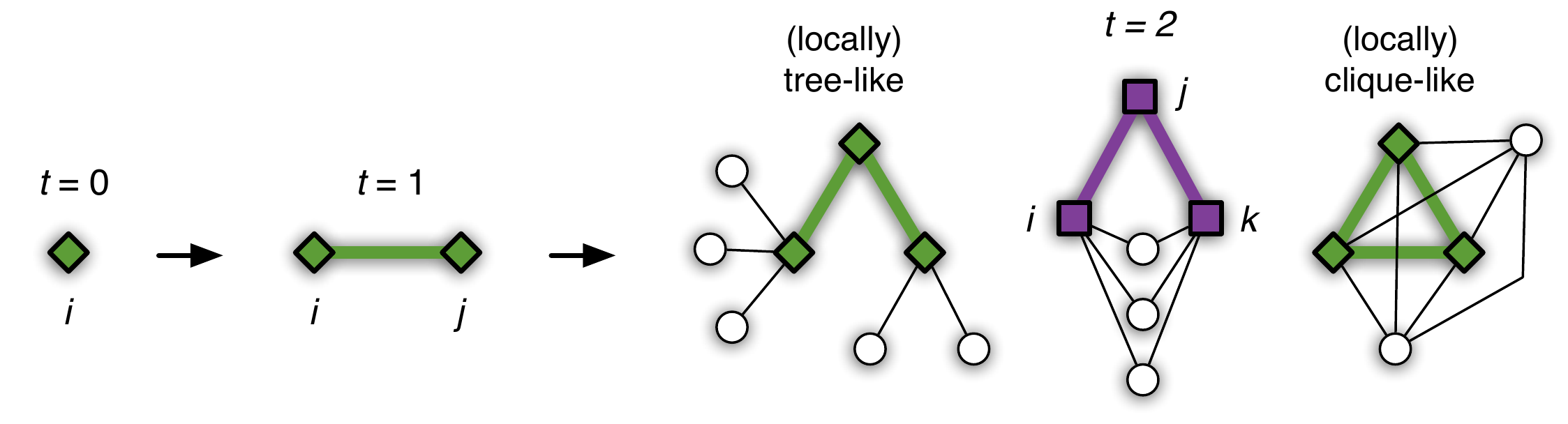}%
	\caption{\label{fig:growth}Growth of convex subgraphs (\diams) in the first two steps $t\leq 2$ of the convex expansion procedure. Notice that non-convex subgraphs (\squrs) indicate the absence of a (locally) tree-like or clique-like structure.}
\end{figure}

\Figref{growth} demonstrates the growth of convex subgraphs in the first two steps $t\leq 2$ of the procedure, $t\in\set{0,\dots,n-1}$. Note that time $t$ measures the number of expansion steps, not including the initialisation step. Initially, the subset $S$ contains a single node $i$, $S=\set{i}$, which is a convex subset. On the first step, a random neighbour $j\in\neigh{i}$ is added, $S=\set{i,j}$, which is still convex. On the next step, another neighbour $k\in\neigh{i}\cup\neigh{j}\setminus\set{i,j}$ is selected, $S=\set{i,j,k}$. In a network that is (locally) tree-like, the subset $S$ induces a convex subtree. Similarly, in a (locally) clique-like network, the subset $S$ induces a triangle and is therefore convex. In any other case, meaning that the subgraph is not a triangle and neither the network is tree-like, the expansion procedure would have to include all the nodes $\neigh{i}\cup\neigh{k}\setminus\set{j}$ in the right side of~\figref{growth} (\squrs) to keep the subset $S$ convex. This may, however, demand additional nodes and so on, possibly resulting in a sudden expansion of the subset $S$. 

The first few steps of the convex expansion procedure thus characterise whether a network is locally either tree-like or clique-like. Furthermore, in the later steps, the absence of any sudden expansion implies that the network is tree-like or clique-like
also globally and therefore convex. A general definition of a fully convex network is a collection of cliques of arbitrary size which are stitched together in a tree, while each pair of cliques can overlap in at most one node~\cite{MS18}. This definition includes a complete graph, where the tree consists of a single node, and also a tree, where all the cliques are single edges.

\begin{figure}[t]
	\centering\includegraphics[width=0.6\textwidth]{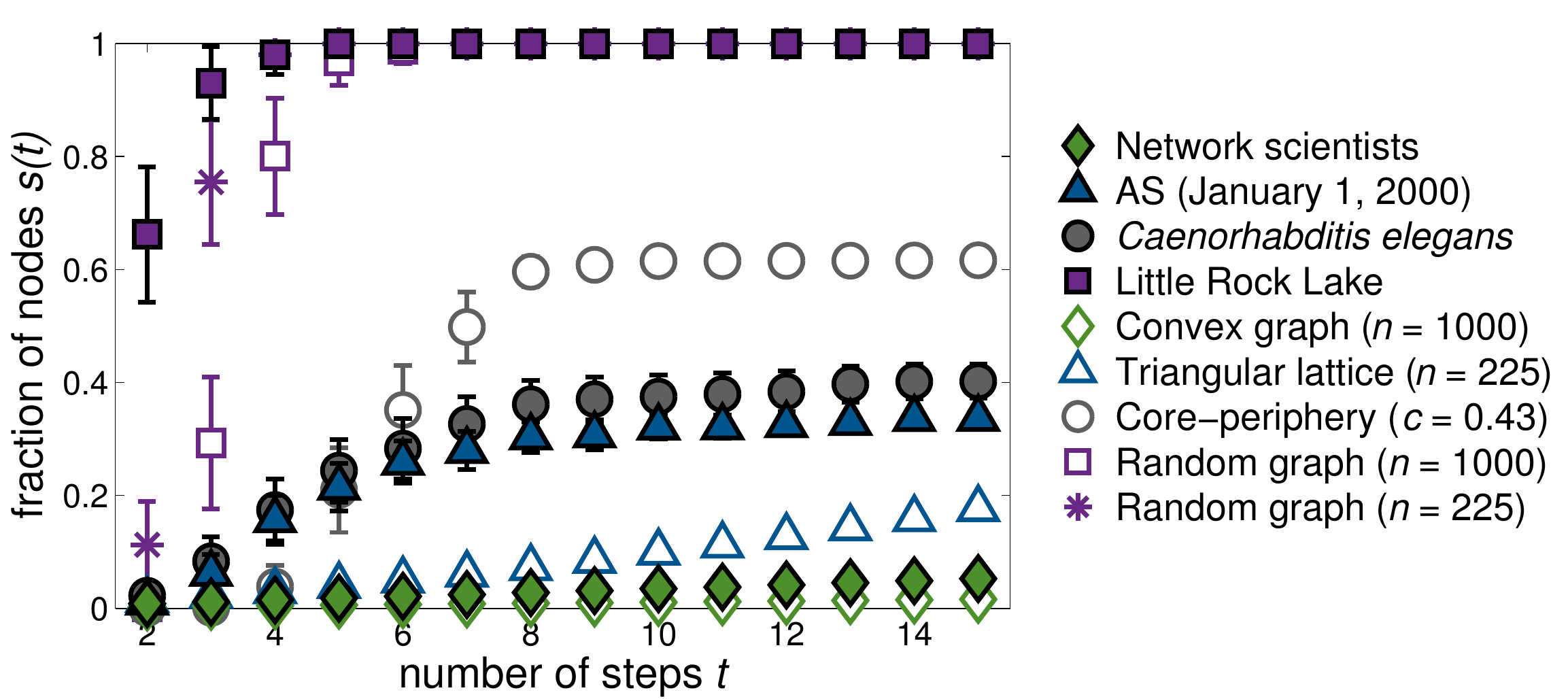}%
	\caption{\label{fig:expansion}Expansion of convex subgraphs in empirical networks and synthetic graphs with $n$ nodes of which $c$ are core nodes in the core-periphery graphs. The plot shows the fraction of nodes $s(t)$ in the growing convex subgraphs at different steps $t\geq 2$, where $s(t)\approx(t+1)/n$ quantifies the presence of a convex structure. The symbols are averages over $100$ runs of convex expansion, while the errors bars show the $99\%$ confidence intervals.}
\end{figure}

Let $s(t)$ denote the average fraction of nodes included in the convex subgraphs or subsets $S$ after $t$ steps of the procedure. According to the adopted time convention, $s(t)\geq (t+1)/n$. In a convex network, $s(t)\approx (t+1)/n$ for any number of steps $t$, whereas $s(t)\gg (t+1)/n$ already for very small $t$ in a non-convex network. As an example, \figref{expansion} shows the evolution of $s(t)$ in different empirical networks and synthetic graphs. The former include social collaboration networks, protein interactions networks, autonomous systems graphs and food webs. The collaboration networks represent jazz musicians~\cite{GD03}, coauthorships between network scientists~\cite{New06a} and Slovenian computer scientists~\cite{BSB12}. The protein interactions networks represent interactions of the \plasm, \scere and \celeg~\cite{SBRBBT06}. The autonomous systems graphs are the Internet maps from 1998, 1999 and 2000~\cite{LKF07}. The food webs represent the species of \littlerock~\cite{Mar91} and Florida Bay~\cite{Pajek}. A thorough description of the empirical networks and the details of the construction of synthetic graphs, together with their basic statistics, are given in~\appref{nets}.

In fully convex graphs that are constructed as random trees with every edge expanded to a clique (\diame in~\figref{expansion}), convex subgraphs grow one node at a time, $s(t)=(t+1)/n$. Similar behaviour is observed for the network scientists coauthorship network. Since the network is a projection of a bipartite graph, it is a union of cliques by construction, while its convex structure further suggests that these are connected together in a tree-like arrangement. On the other hand, there is a sudden expansion of convex subgraphs in the Erd\H{o}s-R\'{e}nyi random graphs~\cite{ER59} having no characteristic structure (\squre in~\figref{expansion}), where $s(t)\approx 1$ after less than five steps $t\leq 5$ in these examples. In fact, the sudden expansion in random graphs occurs at $t=\ln{n}/\ln{\avg{k}}$~\cite{MS18}, where $\avg{k}$ is the expected degree of the nodes. The only empirical network in~\figref{expansion} with a less convex structure than random graphs is the \littlerock food web (\squrs), due to a division of species into trophic levels.

Network convexity can be assessed quantitatively using a measure of convexity $X\in[0,1]$~\cite{MS18}. Let $\Delta s(t)$ denote the average increase in the size of convex subgraphs at step $t$ of the procedure, $\Delta s(t)=s(t)-s(t-1)$. Then,
\begin{eqnarray}
	X & = & 1-\sum_{t=1}^{n-1}\max\set{\Delta s(t)-1/n,\,0}. \label{eq:X}
\end{eqnarray}
Convexity $X$ compares the growth of convex subgraphs $\Delta s(t)$ in a network with the growth in a fully convex graph, where $\Delta s(t)=1/n$ at each step $t$ and $X=1$. In more simple terms, convexity $X$ measures the number of steps of the procedure needed to cover the whole network $t'$ relative to its size $n$, $X=(t'+1)/n$. For instance, convexity of the above coauthorship network is $X=0.85$, while $X\leq 0.03$ for the food web and random graphs. The values of convexity $X$ for all networks and graphs are reported in~\tblref{nets} in~\appref{nets}.

The growth $\Delta s(t)$ shows a notable transition in the autonomous systems graph from 2000 and the \celeg protein interactions network (\trians and \ellps in~\figref{expansion}). Convex subgraphs expand relatively quickly in the first few steps, $\Delta s(t)\gg 1/n$, after which the growth settles, $\Delta s(t)\approx 1/n$. These are both core-periphery networks with a natural division into a densely connected core surrounded by a sparse disconnected periphery~\cite{BE00,Hol05}. In such networks, convex subgraphs quickly cover the network core, which is found to be non-convex, while the growth settles when convex subgraphs start to stretch over a convex periphery~\cite{MS18}.

%
%

\section{\label{sec:random}Convexity under randomisation}

Convexity implies a very specific network structure consisting only of cliques and a tree. Therefore, it is intuitive to expect that convexity is very sensitive to random perturbations of network structure. Indeed, adding a single edge to a tree already creates a cycle, while removing a single edge already destroys a clique. In this section, we study the robustness of convexity in networks and graphs under randomisation by edge rewiring~\cite{MS02}.

All empirical networks analysed in the paper are either connected or reduced to the largest connected component, while all synthetic graphs are almost fully connected. Random rewiring of edges, however, can certainly disconnect the networks and graphs. This presents a problem because the measure of convexity $X$ in~\eqref{X} is only sensible for connected networks~\cite{MS18}. In disconnected networks, there are no (shortest) paths between the nodes of different connected components, thus any argument regarding the inclusion of shortest paths is ill-posed. In fact, the concept of a convex subgraph, which presents the basis of convexity $X$, can only be defined for connected subgraphs, whereas any disconnected subgraph is necessarily non-convex~\cite{HN81,FJ86}.

We here propose a corrected measure of convexity suitable also for disconnected networks that contain one large connected component. Corrected convexity written as $Xs$ is defined as the product $sX$, where $s$ is the fraction of nodes in the largest connected component and convexity $X$ is estimated only from the largest connected component. Hence,
\begin{eqnarray}
	Xs & = & s-s\sum_{t=1}^{sn-1}\max\set{\Delta s(t)-1/sn,\,0} \nonumber\\
	& = & s-\sum_{t=1}^{sn-1}\max\set{s\Delta s(t)-1/n,\,0}, \label{eq:Xs}
\end{eqnarray}
while all other details are the same as in~\eqref{X}. In contrast to convexity $X$, corrected convexity $Xs$ explicitly demands connectivity between the nodes. It is consistent with the definition of a convex subgraph, under which any disconnected subgraph is non-convex, and equivalent to convexity $X$ for connected networks, where $s=1$. Note, however, that the measure is not sensible for disconnected networks with $s\ll 1$.

\begin{figure}[t]
	\centering\includegraphics[width=0.6\textwidth]{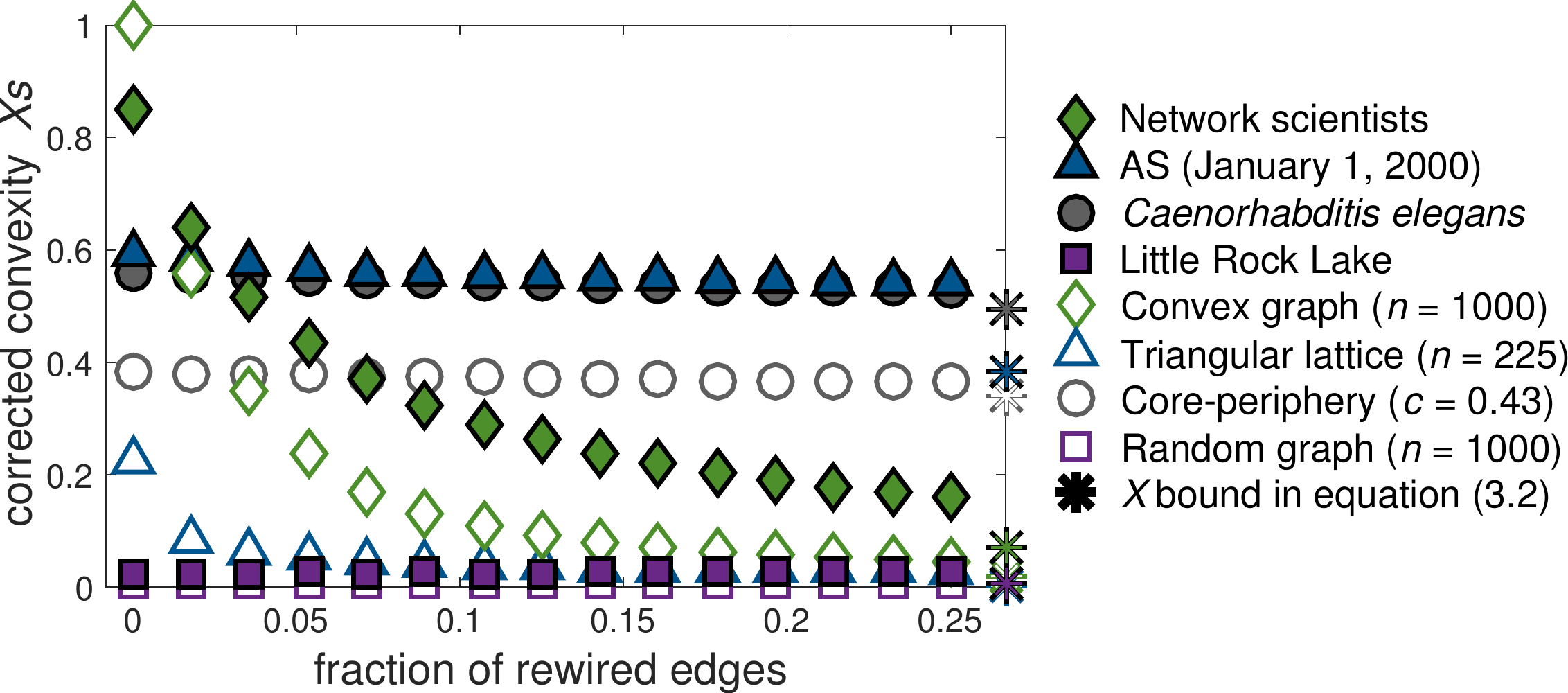}%
	\caption{\label{fig:rewiring}Evolution of convexity in empirical networks and synthetic graphs under degree-preserving randomisation by edge rewiring. The plot shows corrected convexity $Xs$ for different fractions of rewired edges and the asterisks mark the lower bound for $X$. The symbols are averages over $25$ independent randomisations, whereas the errors bars are smaller than the symbol sizes.} 
\end{figure}

\Figref{rewiring} demonstrates the robustness of convexity in different empirical networks and synthetic graphs presented in~\appref{nets}. The plot shows the evolution of corrected convexity $Xs$ under degree-preserving randomisation by edge rewiring~\cite{MS02}, where randomly selected edges are rewired by swapping one of their endpoints. We ensure simple networks and graphs by forbidding any parallel edges or self-edges during the rewiring. At the limit, this process generates a random graph with the same degree sequence as the original network or graph~\cite{NSW01}.

In the case of the \littlerock food web and the Erd\H{o}s-R\'{e}nyi random graphs that lack convexity $Xs\approx 0$ (\squrs and \squre in~\figref{rewiring}), rewiring of edges obviously has no effect. On the other hand, rewiring only a very small fraction of edges already destroys convexity in the network scientists coauthorship network and convex graphs (\diams and \diame in~\figref{rewiring}). In particular, after rewiring $5\%$ of the edges, corrected convexity $Xs$ drops to less than half, although this percentage is about a hundred times smaller than what is normally needed to fully randomise a network or graph~\cite{RPS15}. Thus, as anticipated above, convexity is a very sensitive property that is not resilient even to small perturbations of network structure.

Different behaviour is observed for the autonomous systems graph from 2000 and the \celeg protein interactions network (\trians and \ellps in~\figref{rewiring}), where corrected convexity $Xs$ decreases only moderately. This can be explained by a large number of pendant nodes in these networks, which are nodes with degree one that are connected to the rest of a network by a single edge (see the right side of~\figref{ccore}). Since pendant nodes retain convexity during the convex expansion procedure in~\secref{convex}, and the rewiring process retains the degrees of the nodes, convexity $X$ is bound by
\begin{eqnarray}
	X & \geq & n_1/n, \label{eq:Xn1}
\end{eqnarray}
where $n_1$ is the number of pendant nodes and $n$ the number of all nodes. The bound in~\eqref{Xn1} is marked by the asterisks in~\figref{rewiring} and is non-trivial for scale-free networks~\cite{BA99} due to a large number of nodes with small degree. For example, convexity in the protein interactions network and the corresponding core-periphery graphs (\ellps and \ellpe) can be almost entirely explained by the pendant nodes, while some further dependencies exist in the autonomous systems graph (\trians). In \appref{random}, we analyse the robustness of convexity also under full randomisation, where the bound in~\eqref{Xn1} no longer holds.

%
%

\section{\label{sec:skeleton}Convex skeletons of networks}

\Secref{random} reveals the sensitivity of convexity in networks, which is not resilient even to small random perturbations of network structure. Therefore, any convexity observed in empirical networks in~\secref{convex} might actually be much higher if measured differently, since this fact can be obscured already by a small number of ``random'' edges that decrease convexity. In this section, we investigate this hypothesis by studying the largest high-convexity part of networks obtained by removing the least number of edges, which we call a convex skeleton. 

\subsection{Definition of convex skeleton}

\begin{figure}[t]
	\centering\includegraphics[width=0.4\textwidth,trim={8cm 2cm 8cm 2cm},clip]{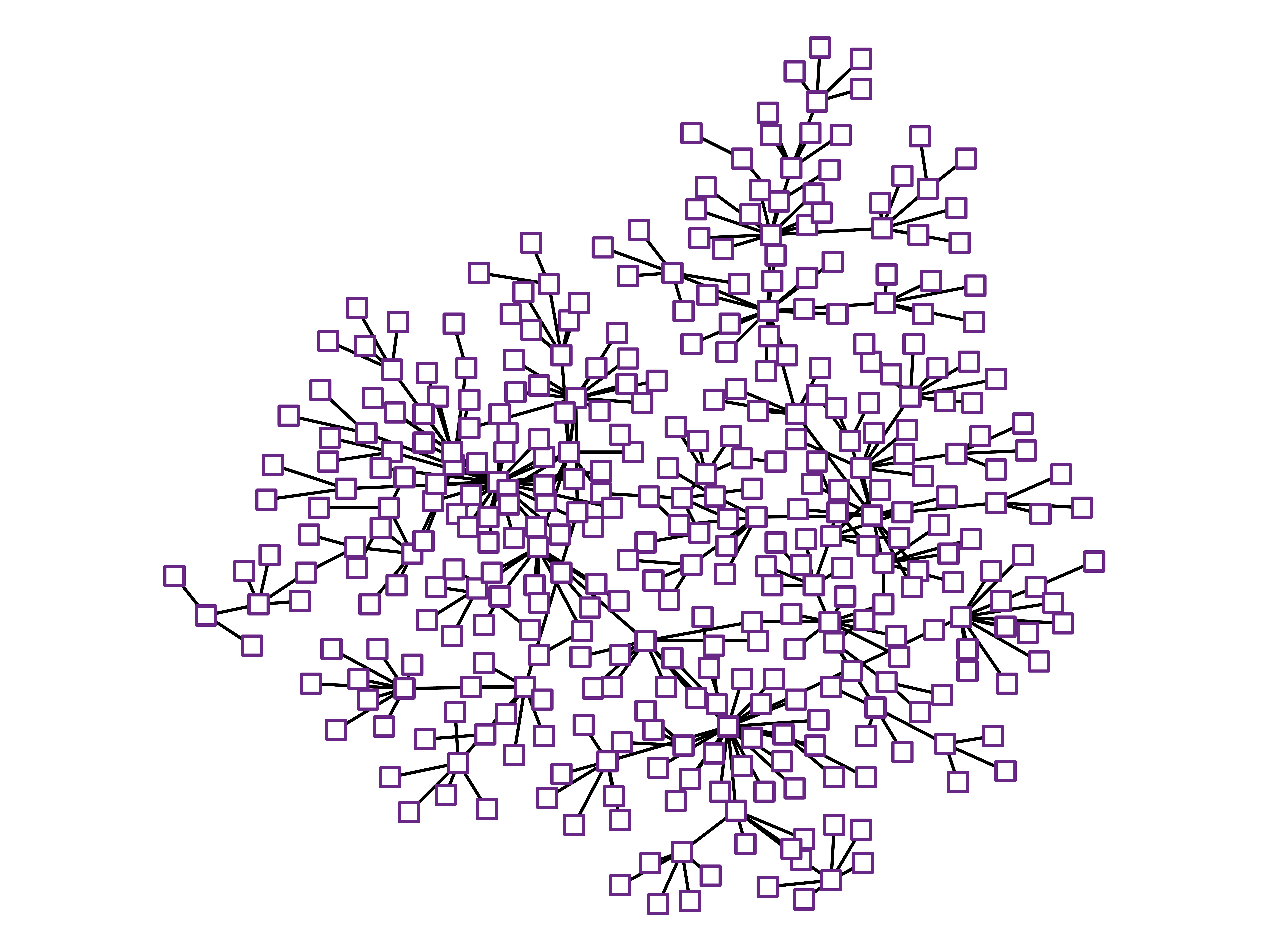}%
	\includegraphics[width=0.4\textwidth,trim={8cm 2cm 8cm 2cm},clip]{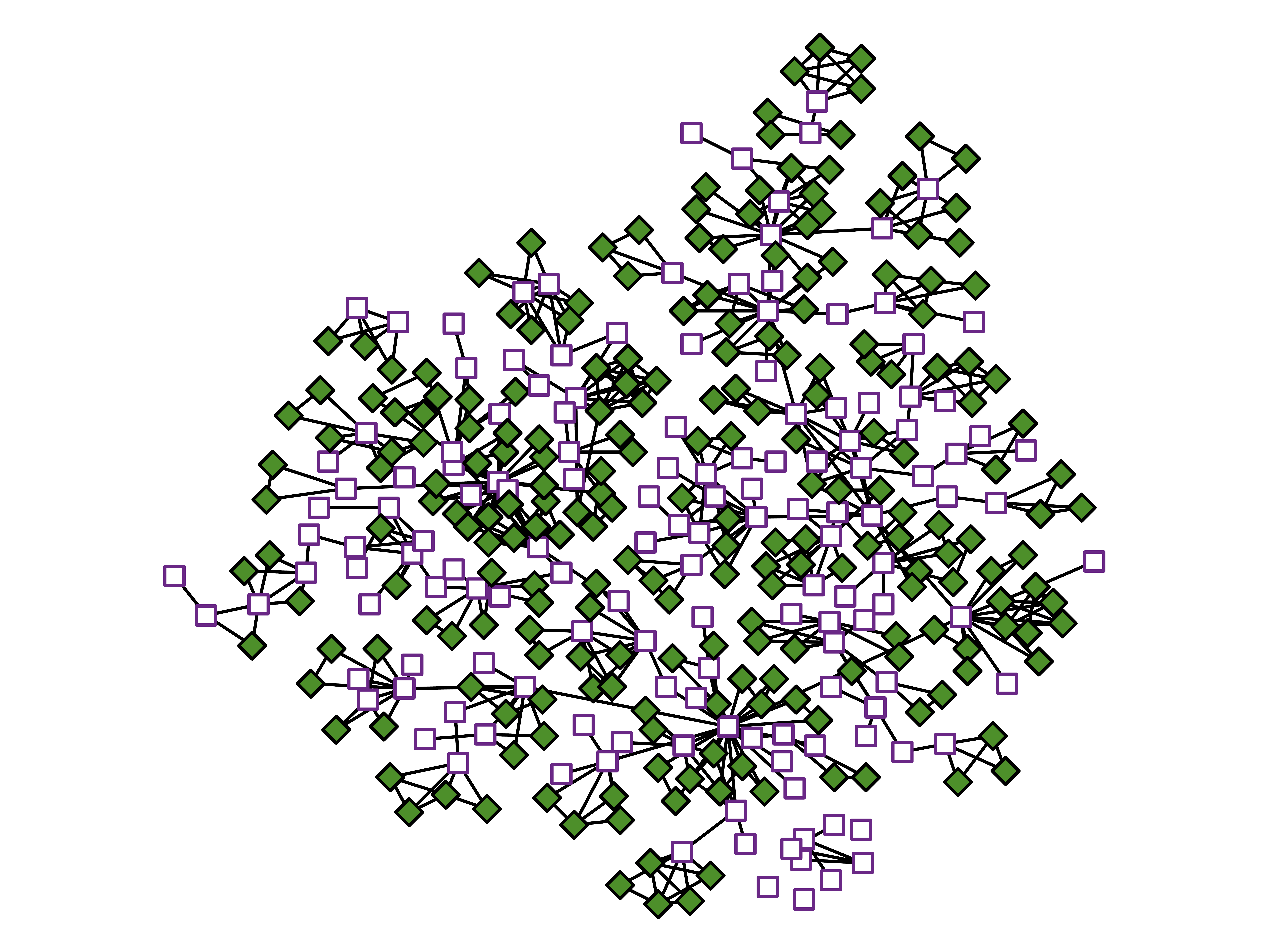}%
	\caption{\label{fig:netsci}(\emph{left})~A spanning tree and (\emph{right})~a convex skeleton of the network scientists coauthorship network extracted as in~\tblref{skeleton}. The nodes with the clustering coefficient above $C>0.5$ are represented with \diams and the remaining nodes with \squre, while the layout was computed with the Large Graph Layout~\cite{ADWM04}.} 
\end{figure}

A convex skeleton is defined as the largest part of a network such that every connected subset of nodes induces a convex subgraph. By definition of a convex network given in~\secref{convex}, a fully convex skeleton would be a collection of cliques glued together in a tree. Each two cliques of a convex skeleton can overlap in at most one node and each cycle is retained in some clique. A convex skeleton is thus a generalisation of a network spanning tree in which each edge can be replaced by a clique of arbitrary size (see the right side of~\figref{netsci}). This implies that every connected network contains at least one convex skeleton, which is a spanning tree, while there are usually many others. Note that we here allow a convex skeleton to contain also nearly cliques with some of the edges missing that are connected in a tree-like manner, whereas any deviation from the ideal structure is measured by corrected convexity $Xs$ defined in~\eqref{Xs}.

Both a spanning tree and a convex skeleton represent a very simple definition of a network backbone~\cite{HLMSW16,CN17} consisting only of cliques and a tree. However, in contrast to a spanning tree, a convex skeleton retains important structural properties of networks by preserving their cliques. For example, \figref{netsci} shows realisations of a spanning tree and a convex skeleton of the network scientists coauthorship network. Consider the average node clustering coefficient $\avg{C}$~\cite{WS98} with the node clustering coefficient defined as $C_i=\frac{2t_i}{k_i(k_i-1)}$, where $t_i$ is the number of triangles including node $i$ and $k_i$ is its degree. Since every coauthorship between three authors is a triangle in a coauthorship network, and larger coauthorships cause larger cliques with many triangles, the network is locally very dense with $\avg{C}=0.74$. This property is retained in a convex skeleton where $\avg{C}=0.75$ on average, while corrected convexity equals $Xs=0.95$. On the other hand, $\avg{C}=0$ and $Xs=1$ for any spanning tree. Furthermore, a convex skeleton consists of more than $90\%$ of the edges in this network, differently from a spanning tree that includes only $41\%$ of the edges.

\subsection{Extraction of convex skeletons}

In what follows, we first present different informed and direct heuristic approaches for extracting convex skeletons from empirical networks by targeted removal of edges. Next, we study the properties of the extracted convex skeletons and show that these retain the degree distribution, clustering, connectivity and also distances between the nodes, while at the same time make the shortest paths between the nodes largely unique. This is because any two nodes in a convex skeleton are either part of the same clique and thus connected by an edge or there exists a unique shortest path traversing the tree.

\begin{figure}[t]
	\centering\includegraphics[width=0.4\textwidth]{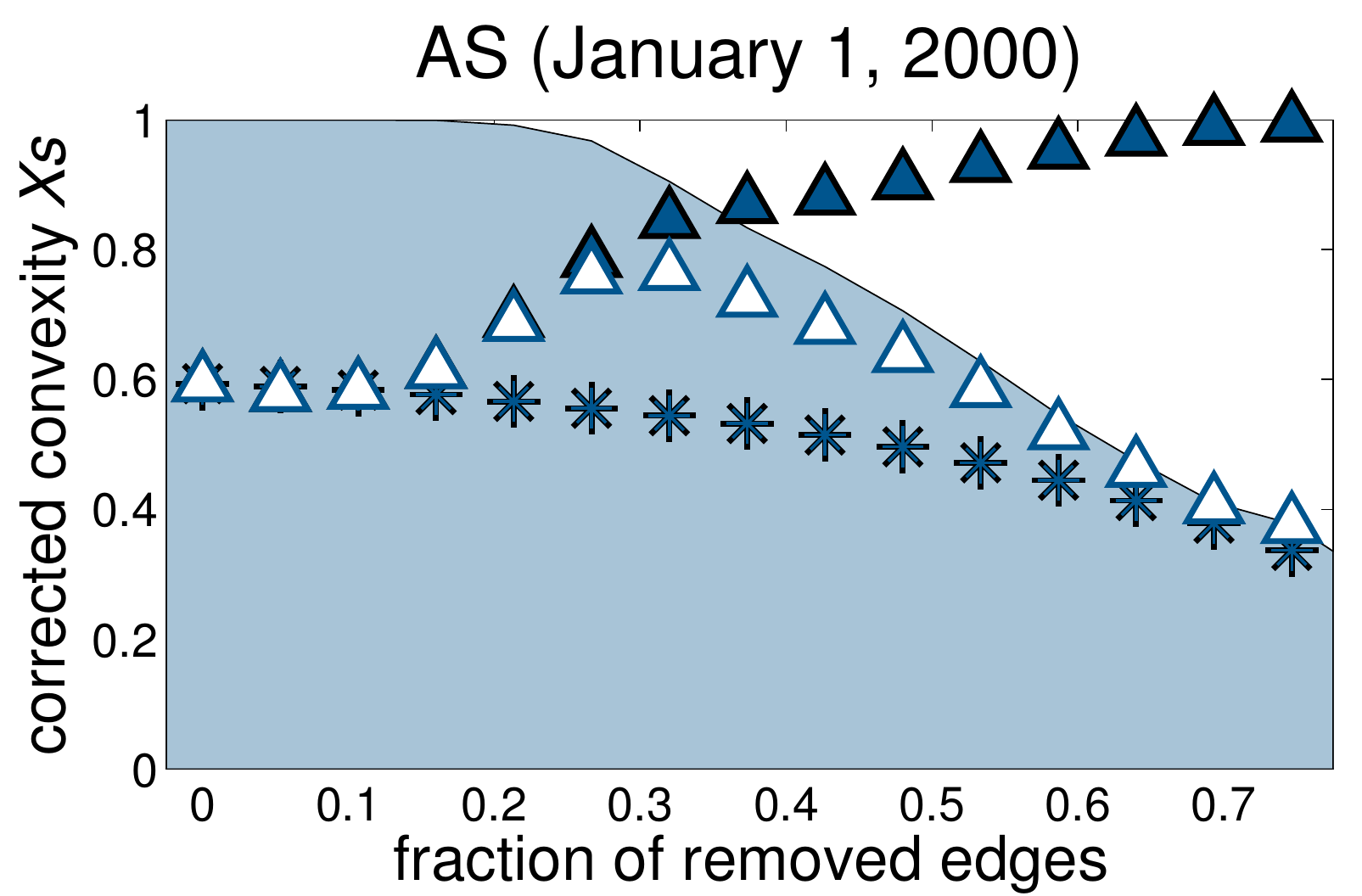}\hskip0.025\textwidth%
	\includegraphics[width=0.4\textwidth]{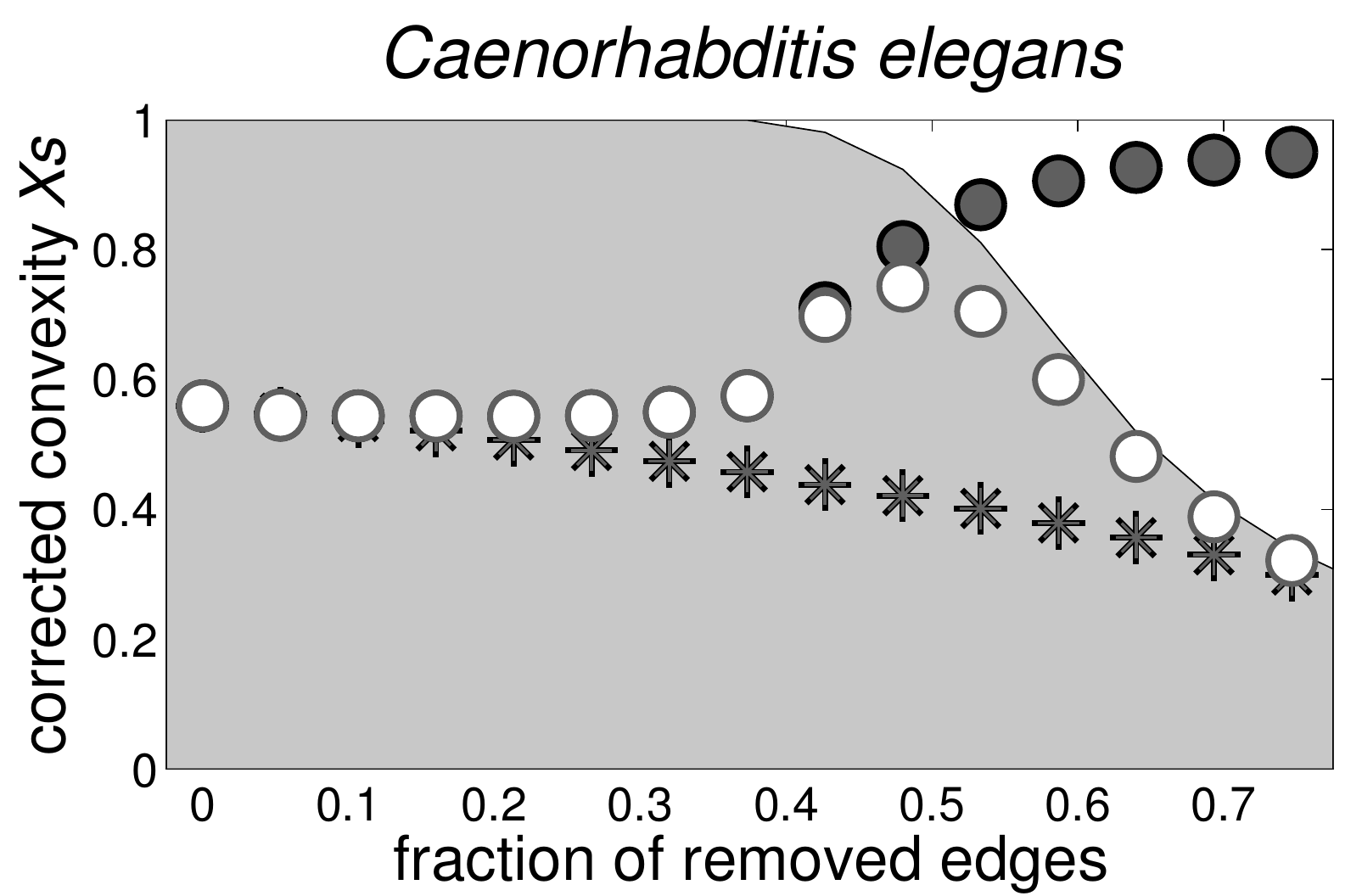}%
	\caption{\label{fig:removal}Evolution of convexity in core-periphery networks under targeted removal of edges based on the node c-centrality. The plots show the fraction of nodes in the largest connected component $s$ (filled areas), convexity $X$ (\trians and \ellps) and corrected convexity $Xs$ (\triane and \ellpe), while the asterisks mark the value of $Xs$ under random removal of edges. The symbols are averages over $25$ independent removals, whereas the errors bars are not visible.} 
\end{figure}

The asterisks in~\figref{removal} show the evolution of corrected convexity $Xs$ in two core-periphery networks under random removal of edges. These are the autonomous systems graph from 2000 and the \celeg protein interactions network. Note that any strategy for removing the edges of a network increases non-corrected convexity $X$ as the network becomes increasingly sparse and thus start to resemble a tree. Yet, the network also starts falling apart decreasing the fraction of nodes in the largest connected component $s$. Under random removal of edges, the networks in~\figref{removal} break apart before convexity $X$ increases, resulting in a monotonically decreasing corrected convexity $Xs$. The same decreasing trend is observed also for targeted removal of edges based on the standard measures of centrality~\cite{Bav50,Fre77,Fre79}.

\begin{figure}[t]
	\centering\includegraphics[width=0.4\textwidth,trim={5cm 2cm 5cm 2cm},clip]{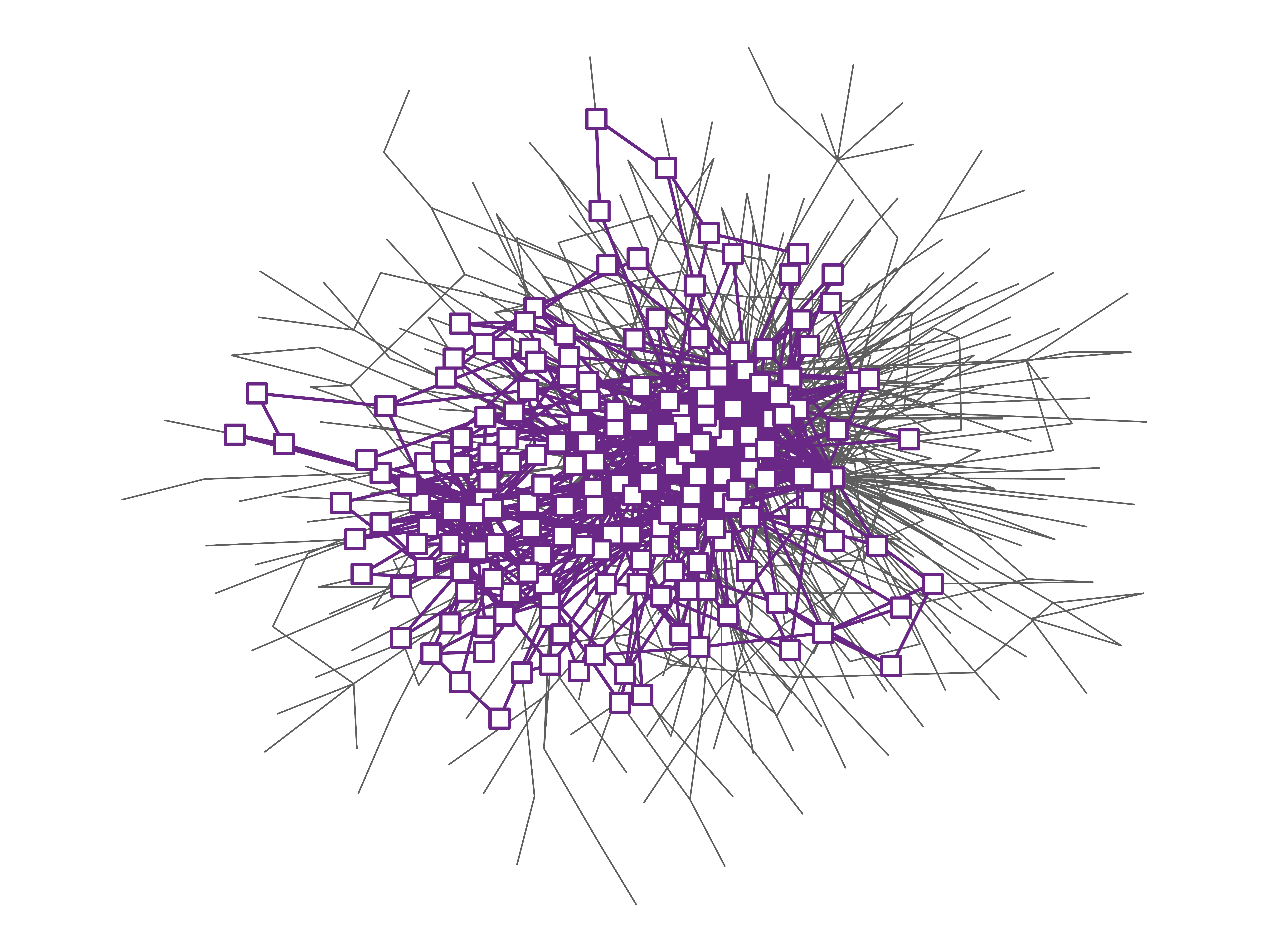}%
	\includegraphics[width=0.4\textwidth,trim={5cm 2cm 5cm 2cm},clip]{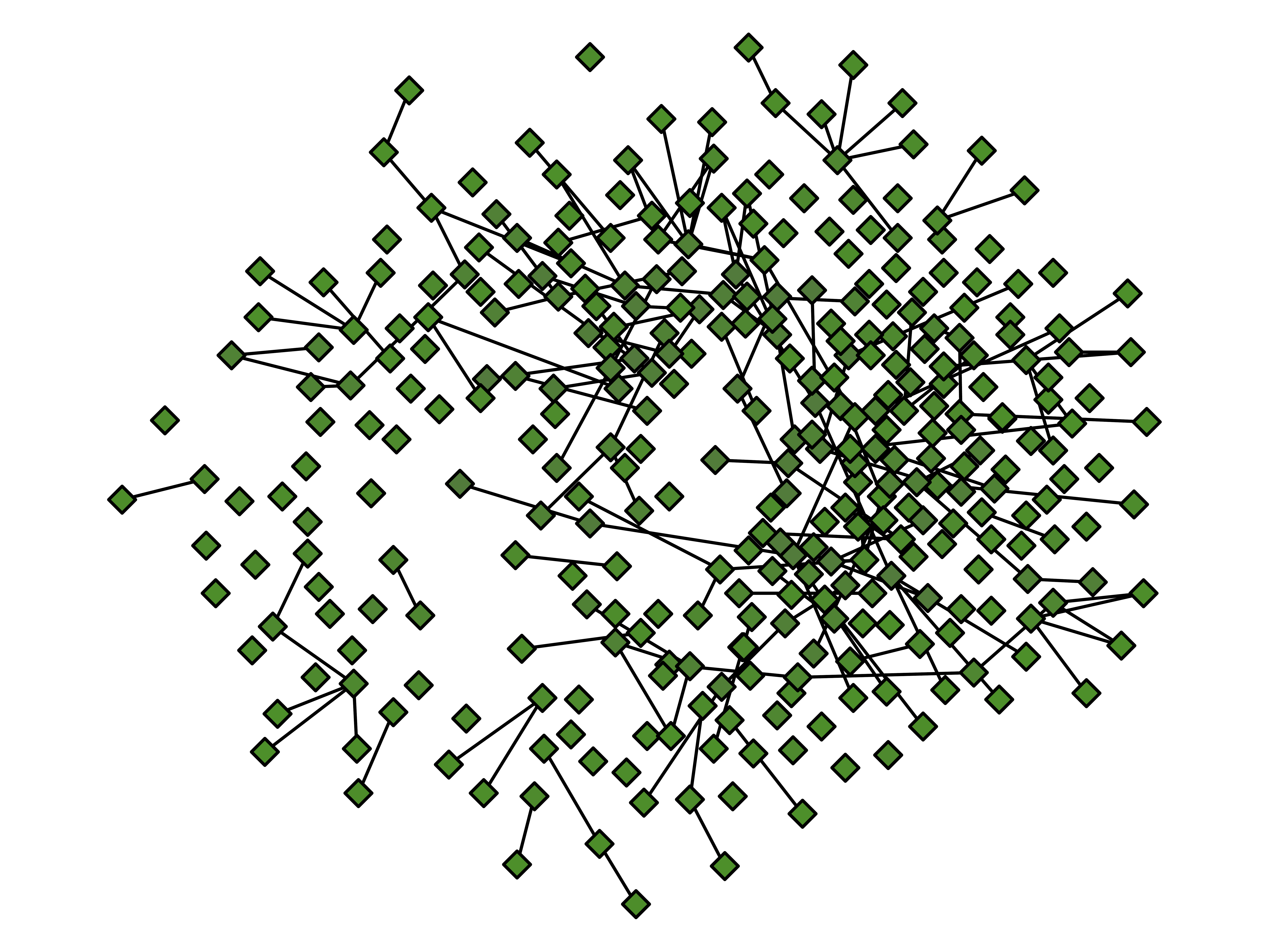}%
	\caption{\label{fig:ccore}Division of the autonomous systems graph from 1999 into (\emph{left})~the c-core represented with \squre and (\emph{right})~a convex periphery represented with \diams. The layout was computed from the full network and is consistent between the figures, while other details are the same as in~\figref{netsci}.} 
\end{figure}

As mentioned in~\secref{convex}, the main source of non-convexity in these networks is the existence of a non-convex core called the network c-core~\cite{MS18}. The c-core is defined as the subset of nodes included in the majority of convex subgraphs after fifteen steps $t=15$ of the convex expansion procedure as in~\figref{expansion}. The particular choice of the threshold $t=15$ does not impact the results as long as $t\geq 2\avg{\ell}$~\cite{MS18}, where $\avg{\ell}$ is the average distance between the nodes. For example, the left side of~\figref{ccore} shows the c-core of the autonomous systems graph from 1999. The removal of edges should therefore target only the edges between the nodes in the c-core, since the remaining periphery is already convex. However, removing the edges between the nodes in the border of the c-core would redirect the shortest paths between the peripheral nodes through the centre of the non-convex c-core, decreasing convexity also in the periphery. Thus, in order to extract a convex skeleton from core-periphery networks, one should target only the edges between the nodes in the centre of the network c-core.

Let $p_i$ be the probability that node $i$ is included in the c-core, which is estimated as the fraction of times node $i$ is included in convex subgraphs after fifteen steps $t=15$ of convex expansion. The distribution of $p$ is extremely bimodal with peaks at $p\approx 0$ and $1$~\cite{MS18}. We define the central nodes of the c-core as those connecting many other nodes in the c-core estimated by $\sum_{j\in\neigh{i}}p_j$, where $\neigh{i}$ is the neighbourhood of node $i$, but not the nodes in the periphery estimated by $\sum_{j\in\neigh{i}}1-p_j$. Hence, the c-core centrality or c-centrality of node $i$ is defined as
\begin{eqnarray}
	c_i & = & \sum_{j\in\neigh{i}}p_j-\sum_{j\in\neigh{i}}1-p_j \nonumber \\
	& = & -k_i+2\sum_{j\in\neigh{i}}p_j, \label{eq:c}
\end{eqnarray}
where $k_i$ is the degree of node $i$. Notice that nodes with high c-centrality are usually not high-degree nodes. Still, in a non-convex network or a random graph where $p_i=1$ for all the nodes, \eqref{c} reduces to $c_i=k_i$.

The symbols in~\figref{removal} show the evolution of convexity under targeted removal of edges based on the node c-centrality. The edges are removed in decreasing order of $c_i+c_j$, where $i$ and $j$ are the endpoints of an edge. We remove a single edge at a time and recalculate the c-centrality of nodes according to~\eqref{c} after each step. Observe that convexity $X$ (\trians and \ellps) increases before the networks start falling apart, resulting in a peak in corrected convexity $Xs$ (\triane and \ellpe). This peak provides a natural way to extract a convex skeleton. For instance, after removing about a third of the edges in the autonomous system graph in the left side of~\figref{removal}, corrected convexity increases from $Xs=0.59$ to $0.77$, when $X=0.85$ and $s=0.91$. Moreover, as we show below, the network actually contains a convex skeleton with $Xs=0.90$ that retains more than $75\%$ of its edges. 

In the remaining, we consider also approaches for targeted removal of edges that try to extract a convex skeleton more directly. However, directly optimising corrected convexity $Xs$ in a network comes with a number of problems. First, the computation of convexity $X$ is inevitably quadratic operation with $\cmp{nm}$~\cite{convex}, where $n$ is the number of nodes and $m$ the number of edges. The complexity of removal of edges would therefore be at least cubic $\cmp{nm^2}$ and applicable only to very small networks. Second, random variations in the estimation of convexity $X$ are often larger than the changes resulting from the removal of a single edge. Third, optimising corrected convexity $Xs$ does not guarantee a convex skeleton that retains important properties of a network, since any network spanning tree is also convex with $Xs=1$.

A better approach is to directly optimise the properties of a network that should ideally be retained in a convex skeleton such as the average node clustering coefficient $\avg{C}$~\cite{skeleton}. Consider an edge with endpoints $i$ and $j$, and let $\Delta C_i$ denote the change in the clustering coefficient of node $i$ after the removal of the mentioned edge. In order to extract a convex skeleton, we remove the edges from a network in a decreasing order of $\Delta C_i+\Delta C_j$, while ensuring that the network stays connected. Nevertheless, since we remove between $0.1\%$ and $1\%$ of edges at a time from the networks below, while we ensure the connectivity only for the removal of individual edges, these can become disconnected.

\begin{figure}[t]
	\centering\includegraphics[width=0.285\textwidth]{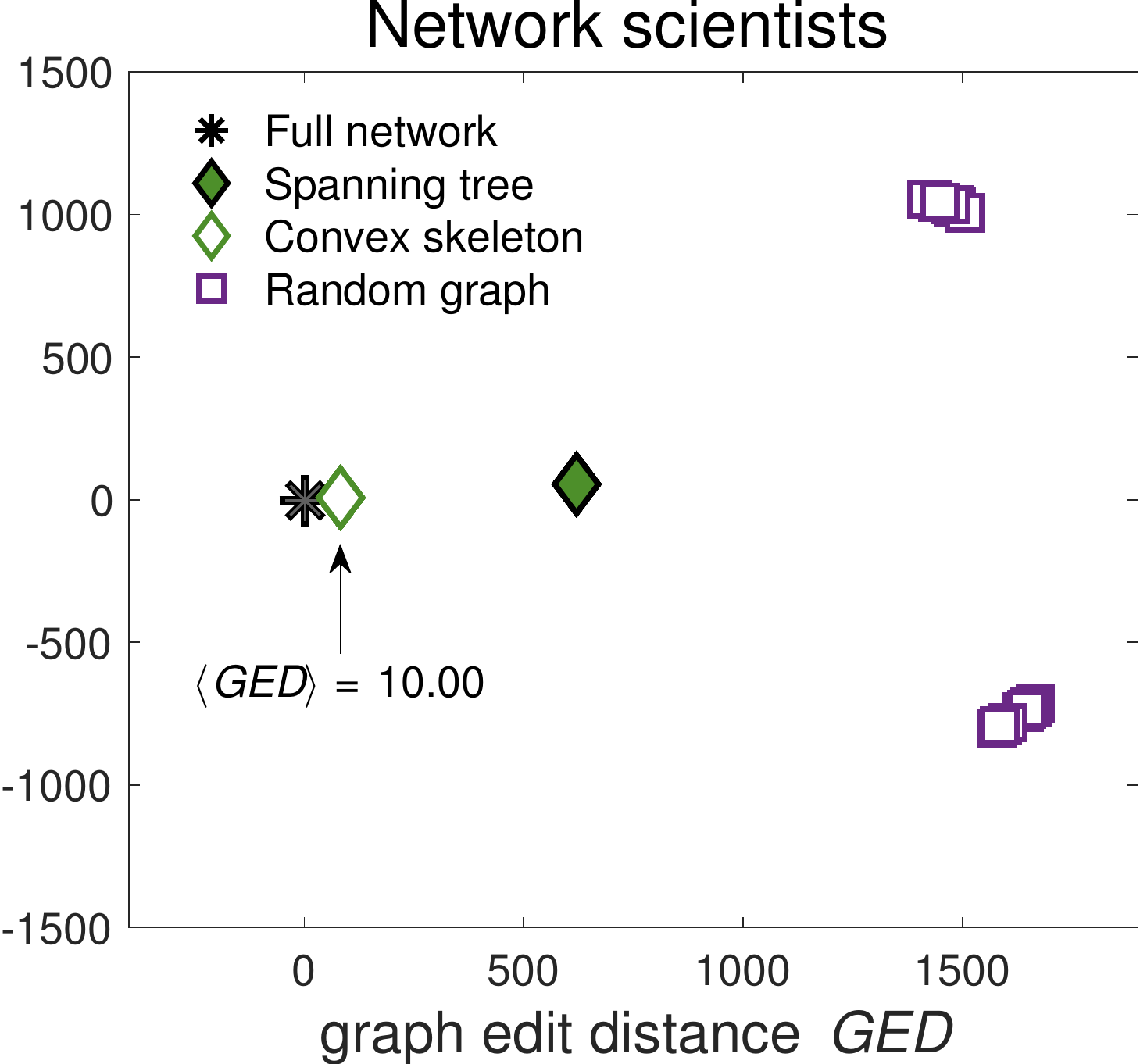}\hskip0.05\textwidth%
	\includegraphics[width=0.285\textwidth]{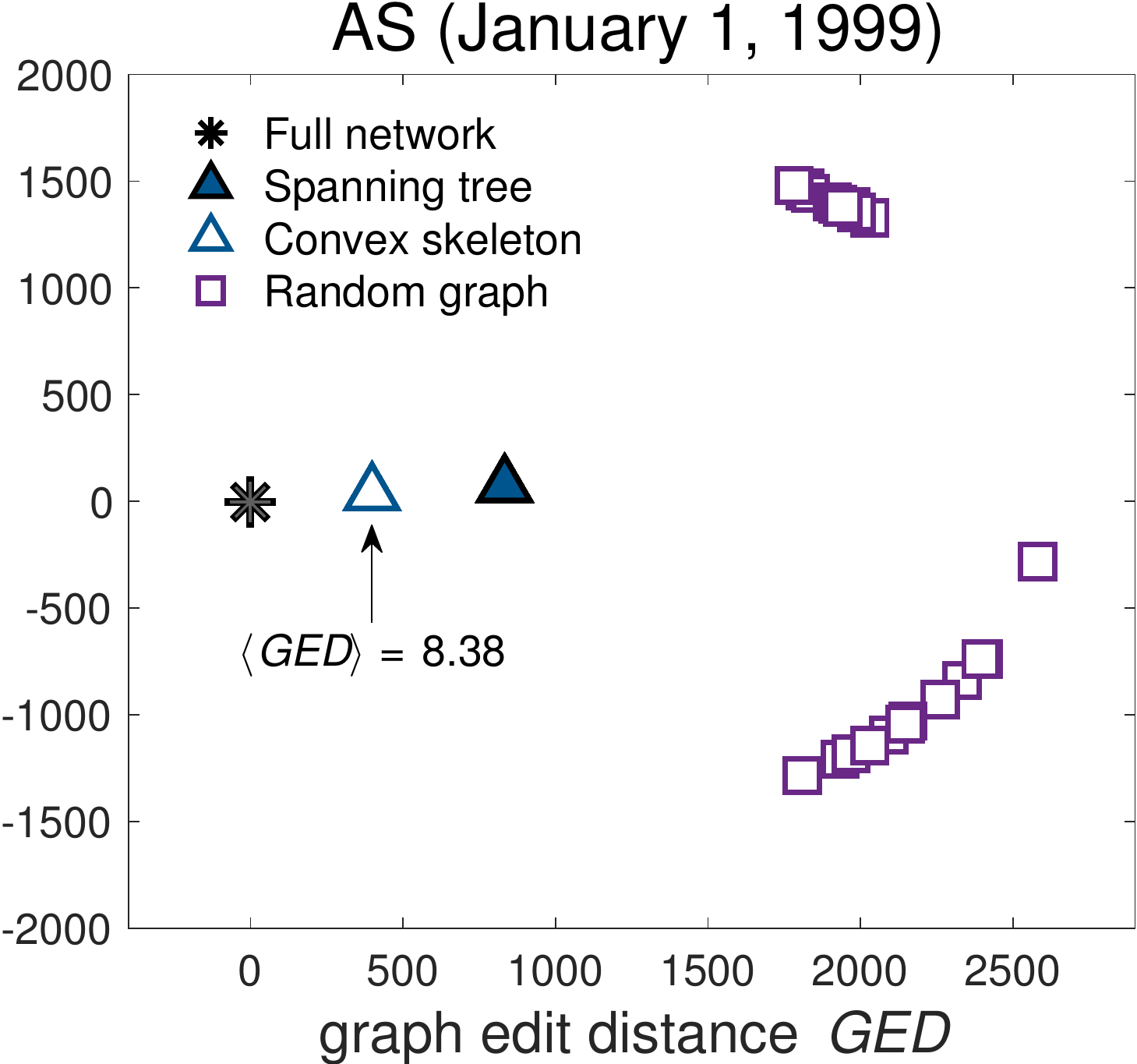}\hskip0.05\textwidth%
	\includegraphics[width=0.285\textwidth]{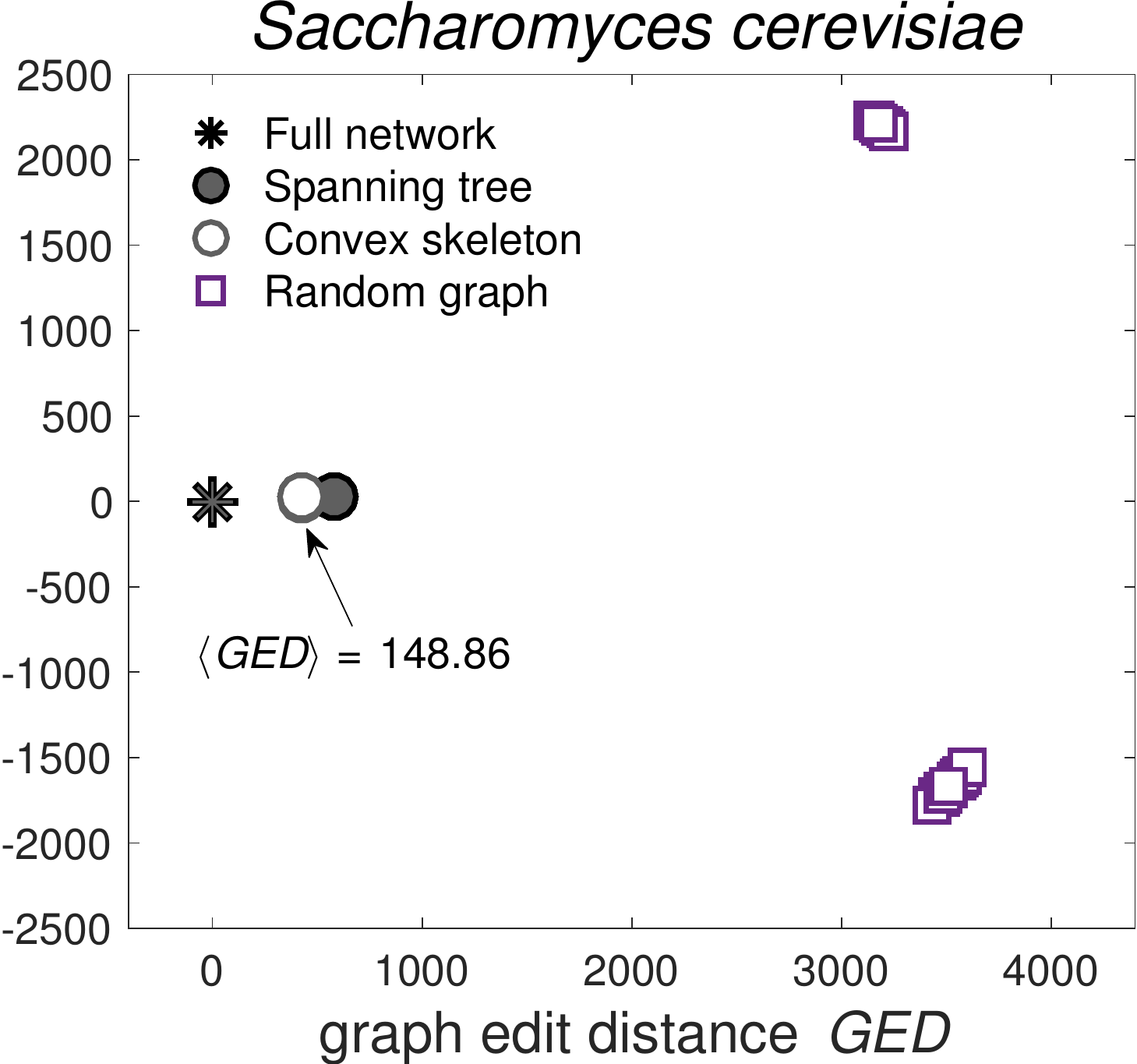}%
	\caption{\label{fig:mds}Multidimensional scaling maps of empirical networks, random graphs, spanning trees and convex skeletons extracted as in~\tblref{skeleton}. The maps are performed over graph edit distance {\it GED} between $25$ independent realisations and have been translated thus the networks appear in the origin. Note that the symbols for spanning trees and convex skeletons overlap. The annotations show the average distance $\langle${\it GED}$\rangle$ between the convex skeletons.} 
\end{figure}

The extraction of convex skeletons is obviously non-deterministic. To see this, consider a ring graph, where a convex skeleton can be obtained by removing an arbitrary edge. Convex skeletons extracted from empirical networks are still remarkably deterministic. \Figref{mds} shows multidimensional scaling maps of different empirical networks, the corresponding Erd\H{o}s-R\'{e}nyi random graphs~\cite{ER59}, network spanning trees and convex skeletons. The empirical networks include the network scientists coauthorship network, the autonomous systems graph from 1999 and the \scere protein interactions network. The maps are performed over graph edit distance {\it GED} defined as the number of edges that need to be inserted or deleted to translate one graph or network to another.

The maps show that convex skeletons are much closer to empirical networks than spanning trees and random graphs. More importantly, the extracted convex skeletons are very similar to one another. The average distance $\langle${\it GED}$\rangle$ between the convex skeletons is only $1.20\%$ of the edges for the coauthorship network, $0.99\%$ for the autonomous systems graph and $9.45\%$ for the protein interactions network. The particular structure of empirical networks therefore allows for a robust extraction of convex skeletons, in contrast to say a ring graph. As shown in the remaining of the paper, convex skeletons also retain the node degree distribution, network clustering and connectivity, distances between the nodes, node position and network community structure.

In a more broader context, removal of network edges as for extracting convex skeletons is a popular technique for analysing the structure and dynamics of empirical networks. For instance, the emergence of connectivity in a network can be studied by random removal of edges known as inverse bond percolation in physics literature~\cite{Bar16}. Edges that destroy the connectivity of a network are called bridges~\cite{WTL18}. Moreover, random and targeted removal of edges is also commonly adopted for network sampling~\cite{BSB17}, sparsification or simplification~\cite{CN17} and for revealing network hierarchy~\cite{Cos18}. Yet another example is removal of spurious edges in a network while retaining its basic properties~\cite{ZC12} or highlighting specific properties~\cite{VSKLBLW11}.

\subsection{Properties of convex skeletons}

\Tblref{skeleton} shows selected statistics of convex skeletons extracted from different social collaboration and protein interactions networks, autonomous systems graphs and food webs. The details of the networks are given in~\appref{nets}. For an easier comparison, \tblref{skeleton} also shows the statistics of network spanning trees~\cite{backbone}, although all these directly follow from the definition of a tree.

\begin{table}[p]
	\caption{\label{tbl:skeleton}Statistics of empirical networks (N), spanning trees (ST) and convex skeletons (CS) extracted by targeted removal of edges based on the node clustering coefficient (we remove $1\%$ of edges at a time or $0.1\%$ for protein interactions networks). The statistics show the average node degree $\avg{k}$ and clustering coefficient $\avg{C}$, the average number of shortest paths or geodesics $\avg{\sigma}$ and corrected convexity $Xs$. The values are averages over $25$ independent realisations and $100$ runs of convex expansion.}
	\begin{sideways} \begin{tabular}{llrrccccrccccc}
		\multirow{2}{*}{Class} & \multirow{2}{*}{Network} & \multicolumn{3}{c}{degree $\avg{k}$} & \multicolumn{3}{c}{clustering $\avg{C}$} & \multicolumn{3}{c}{geodesics $\avg{\sigma}$} & \multicolumn{3}{c}{convexity $Xs$} \\
		& & \ncsst & \ncsst & \ncsst & \ncsst \\\hline
		\multirow{3}{*}{Collaboration} & \jazz & $27.70$ & $11.06$ & $1.99$ & $0.62$ & $0.81$ & $0.00$ & $9.71$ & $1.97$ & $1.00$ & $0.12$ & $0.84$ & $1.00$ \\ 
		& \netsci & $4.82$ & $4.40$ & $1.99$ & $0.74$ & $0.75$ & $0.00$ & $2.66$ & $1.47$ & $1.00$ & $0.85$ & $0.95$ & $1.00$ \\ 
		& \cmpsci & $4.75$ & $3.25$ & $1.99$ & $0.48$ & $0.54$ & $0.00$ & $4.08$ & $1.42$ & $1.00$ & $0.64$ & $0.95$ & $1.00$ \\\hline 
		\multirow{3}{*}{\shortstack[l]{Protein\\ interactions}} & \plasm & $4.15$ & $2.26$ & $2.00$ & $0.02$ & $0.07$ & $0.00$ & $3.71$ & $1.77$ & $1.00$ & $0.43$ & $0.95$ & $1.00$ \\ 
		& \scere & $2.67$ & $2.16$ & $2.00$ & $0.07$ & $0.10$ & $0.00$ & $2.58$ & $1.19$ & $1.00$ & $0.68$ & $0.88$ & $1.00$ \\ 
		& \celeg & $4.14$ & $2.43$ & $2.00$ & $0.06$ & $0.12$ & $0.00$ & $6.79$ & $3.03$ & $1.00$ & $0.56$ & $0.85$ & $1.00$ \\\hline 
		\multirow{3}{*}{\shortstack[l]{Autonomous\\systems}} & \oreg{1998} & $3.50$ & $2.70$ & $2.00$ & $0.18$ & $0.21$ & $0.00$ & $3.87$ & $2.32$ & $1.00$ & $0.66$ & $0.91$ & $1.00$ \\ 
		& \oreg{1999} & $4.58$ & $3.18$ & $2.00$ & $0.18$ & $0.27$ & $0.00$ & $3.54$ & $2.05$ & $1.00$ & $0.49$ & $0.95$ & $1.00$ \\ 
		& \oreg{2000} & $3.94$ & $2.96$ & $2.00$ & $0.20$ & $0.25$ & $0.00$ & $4.81$ & $3.07$ & $1.00$ & $0.59$ & $0.90$ & $1.00$ \\\hline 
		\multirow{3}{*}{Food webs} & \littlerock & $26.60$ & $4.31$ & $1.99$ & $0.32$ & $0.69$ & $0.00$ & $22.13$ & $4.32$ & $1.00$ & $0.02$ & $0.82$ & $1.00$ \\ 
		& \baywet & $32.42$ & $3.98$ & $1.98$ & $0.33$ & $0.79$ & $0.00$ & $9.17$ & $1.37$ & $1.00$ & $0.03$ & $0.92$ & $1.00$ \\ 
		& \baydry & $32.91$ & $5.02$ & $1.98$ & $0.33$ & $0.82$ & $0.00$ & $9.37$ & $1.65$ & $1.00$ & $0.03$ & $0.93$ & $1.00$ 
	\end{tabular} \end{sideways}
\end{table}

The average corrected convexity of the extracted convex skeletons over all networks is $\avg{Xs}=0.90$, which can be considered as high-convexity parts of these networks. Besides, this implies that the skeletons are almost fully connected as the fraction of nodes in the largest connected component is $s\geq Xs$. We stress that the adopted process of edge removal does not optimise convexity directly, thus a convex skeleton is an emerging property of these networks. The fact might be particularly interesting as the removal process is almost the exact opposite of small-world network models~\cite{WS98,NW99b}, where the edges of a ring lattice are randomly rewired or new random edges are added. In contrast to a ring lattice, a convex skeleton suggest a tree-like structure. This has important implications for the distances between the nodes. While the average distance $\avg{\ell}$ in a ring on $n$ nodes is in $\cmp{n}$~\cite{New10}, $\avg{\ell}=\cmp{\sqrt{n}}$ in a random tree~\cite{MM70} and $\avg{\ell}=\cmp{\ln{n}}$ in a balanced tree, consistently with small-world networks.

The average node clustering coefficient $\avg{C}$ in the extracted convex skeletons is expectedly higher than in the full networks, although obviously no new triangles are created. Furthermore, the shortest paths or geodesics between the nodes become largely unique, with the average number of geodesics $\avg{\sigma}<2$ in most cases. At the same time, the average distance between the nodes $\avg{\ell}$ remains comparable, due to an underlying tree-like structure in a convex skeleton. A convex skeleton might thus represent a suitable model for studying different dynamic processes on networks such as navigation and synchronisation~\cite{BBV08}.

\Tblref{skeleton} also shows the average node degree $\avg{k}$, which allows to estimate the fraction of edges retained in the convex skeletons. On the one hand, this fraction is larger than $90\%$ in the network scientists coauthorship network, while $70$-$75\%$ in the computer scientists coauthorship network and the autonomous systems graphs. On the other hand, one has to remove $85$-$90\%$ of the edges in the food webs to reveal a convex skeleton. This can be anticipated from a large redundancy of geodesics in food webs needed for an ecosystem to survive. For instance, the average number of geodesics in the \littlerock food web is $\avg{\sigma}>20$. A convex skeleton therefore does not represent a very sensible abstraction of food webs.

\begin{figure}[t]
	\centering\includegraphics[width=0.35\textwidth]{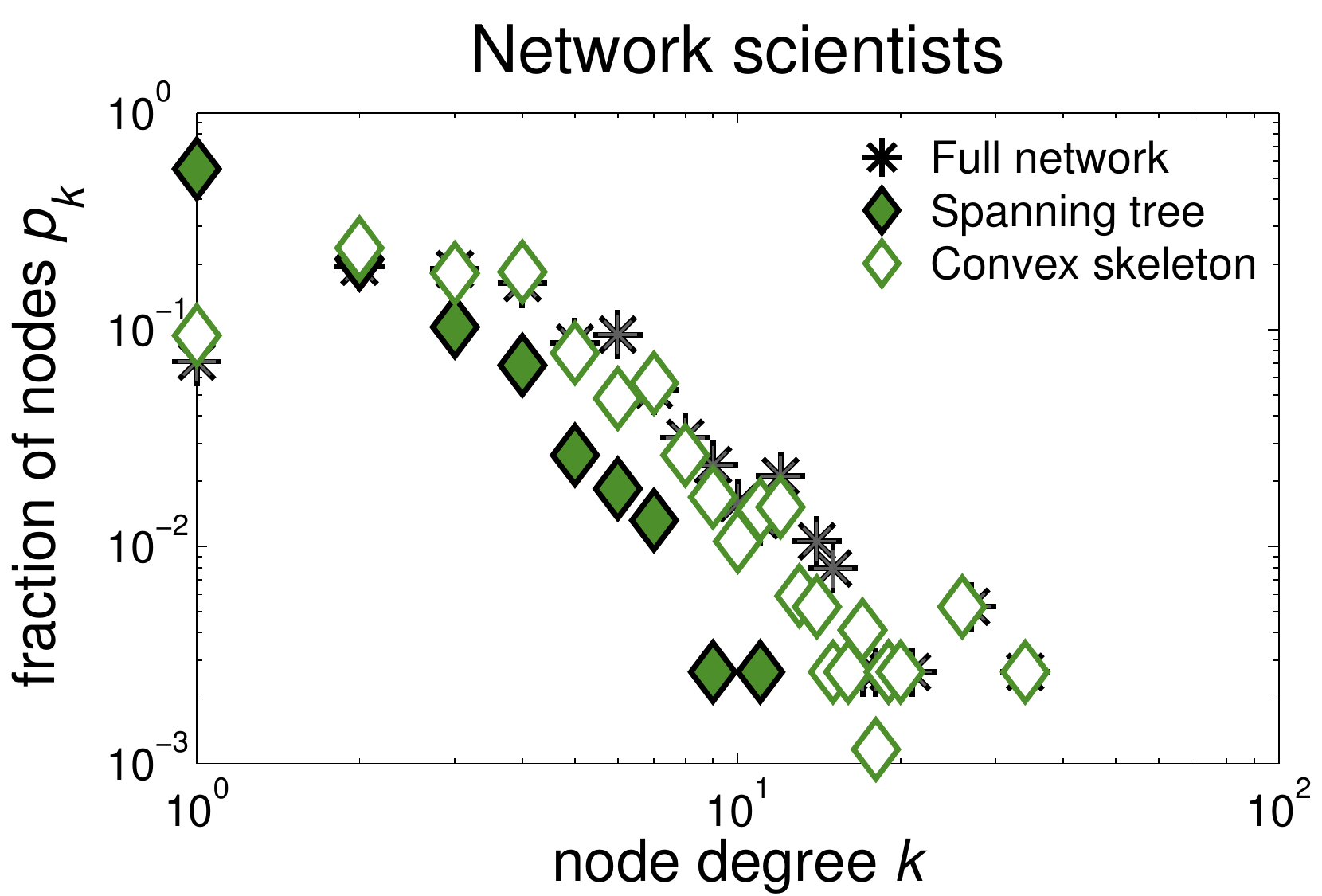}\hskip0.025\textwidth%
	\includegraphics[width=0.35\textwidth]{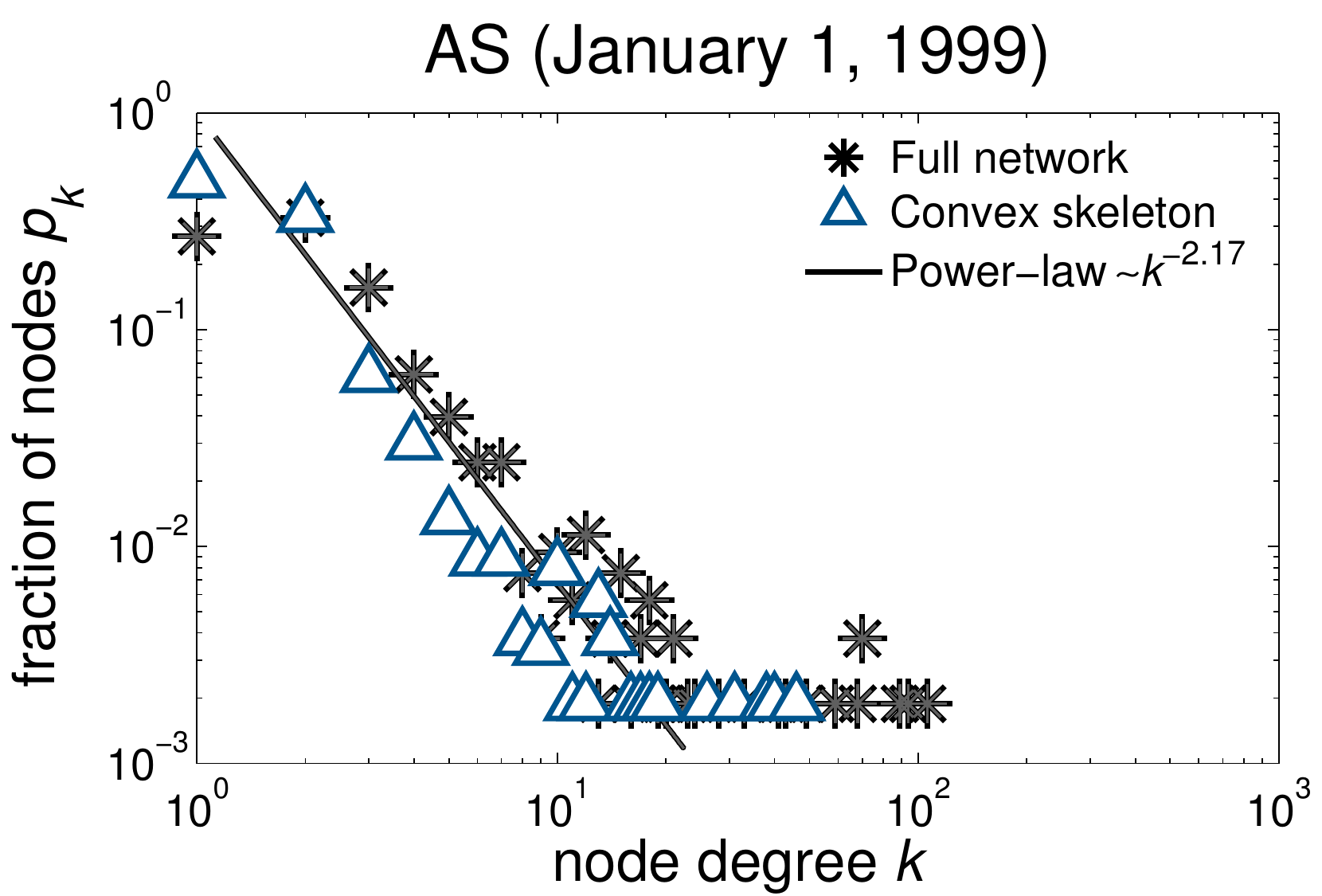}\\\vskip0.0125\textwidth%
	\includegraphics[width=0.35\textwidth]{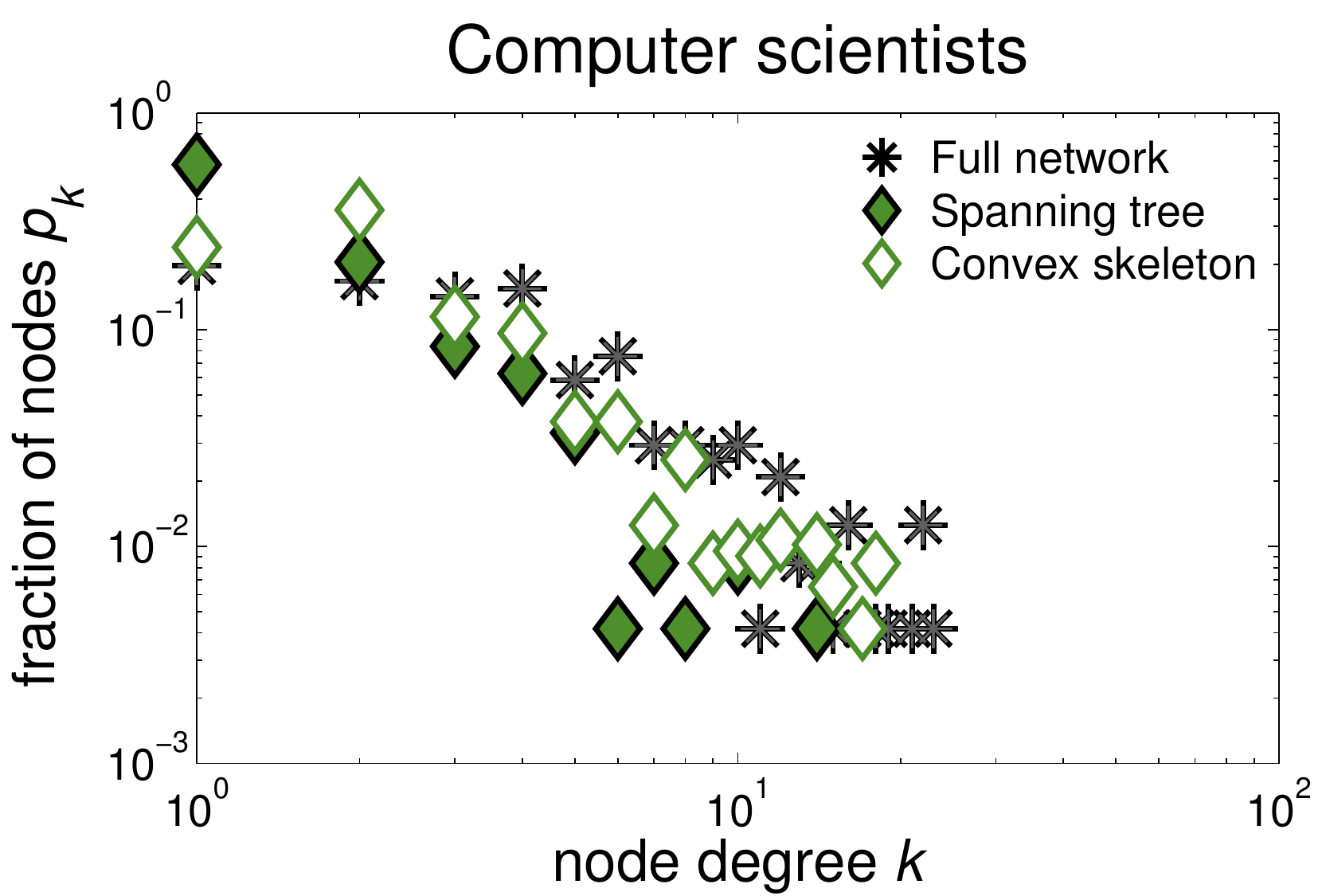}\hskip0.025\textwidth%
	\includegraphics[width=0.35\textwidth]{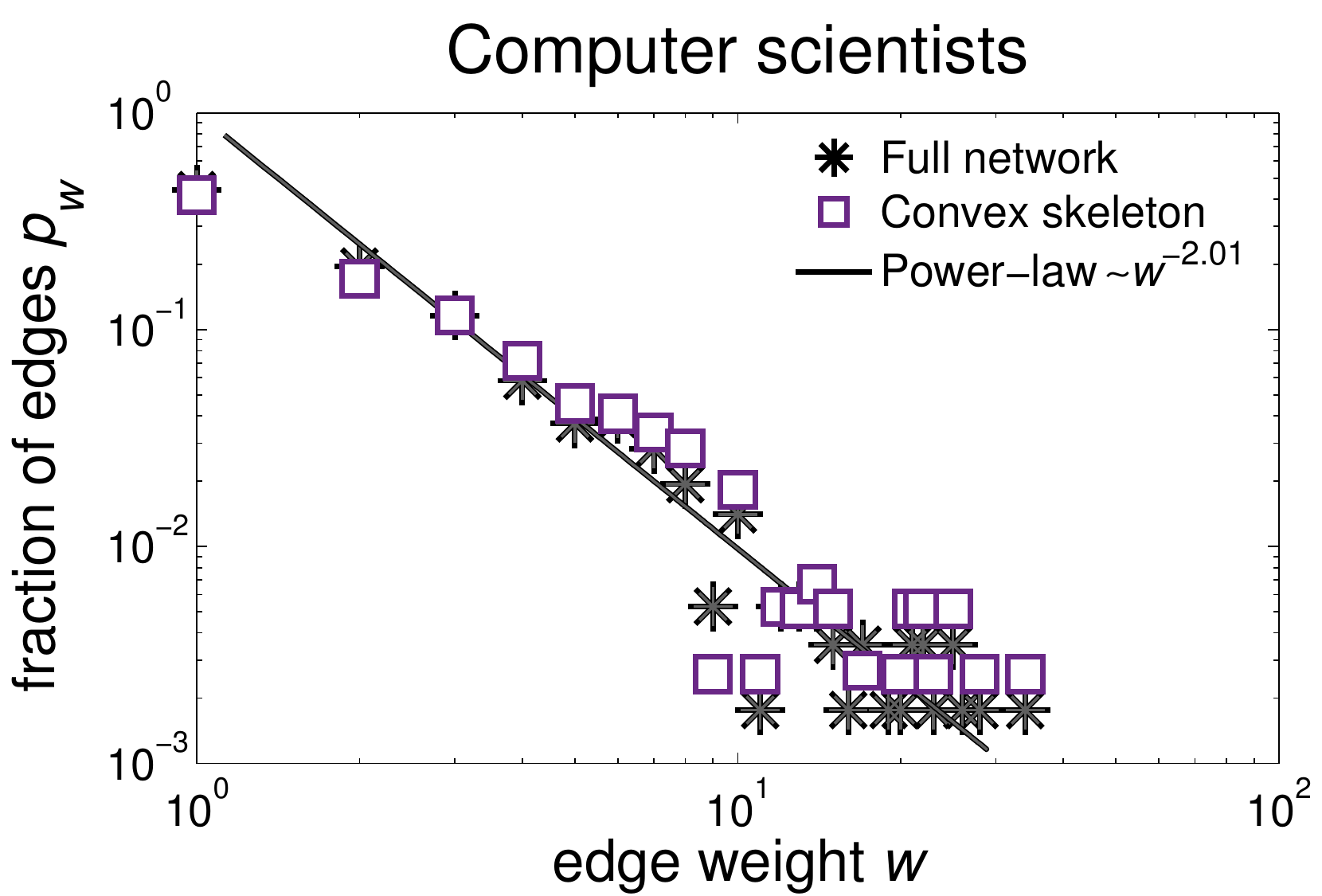}\\\vskip0.0125\textwidth%
	\includegraphics[width=0.35\textwidth]{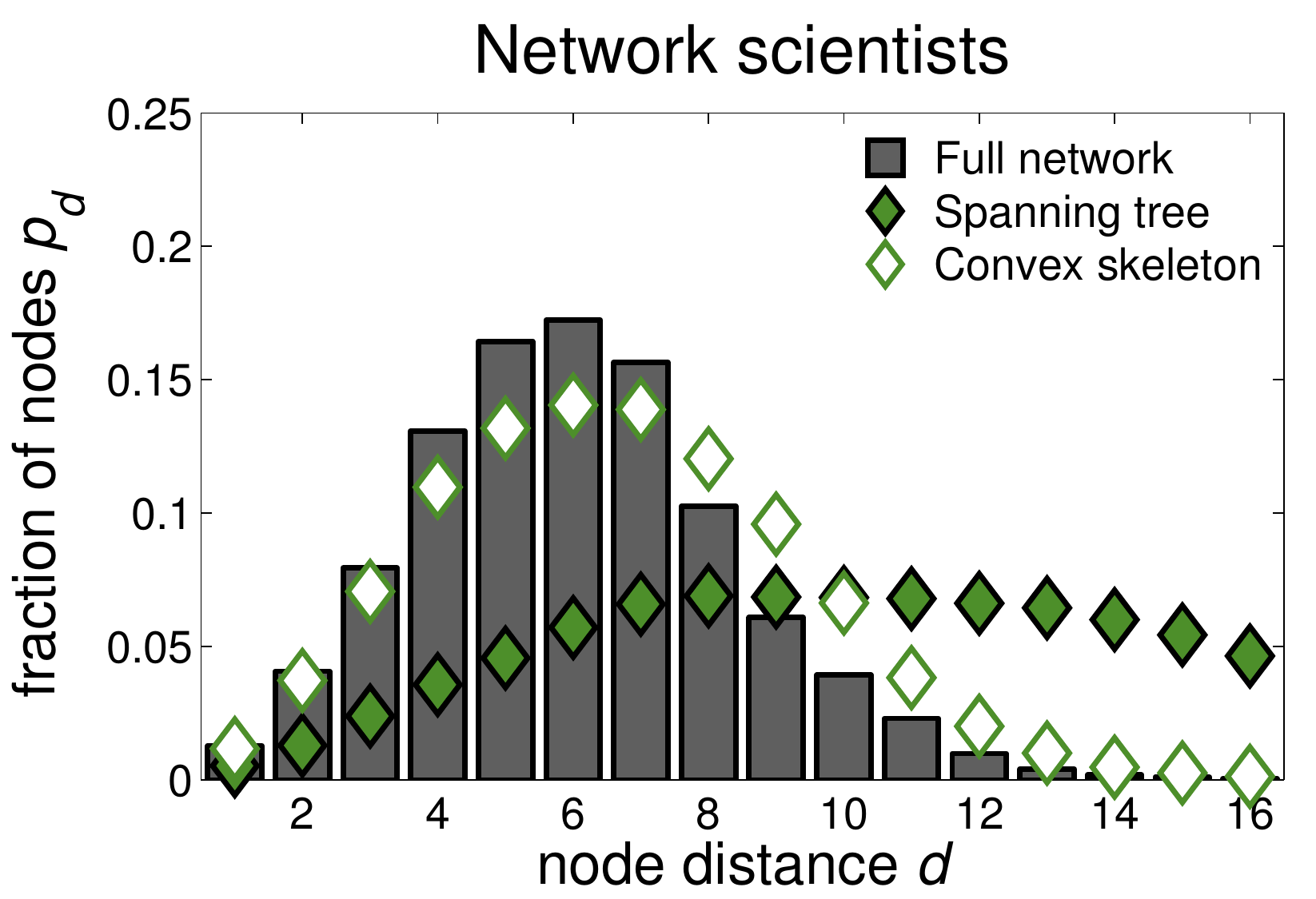}\hskip0.025\textwidth%
	\includegraphics[width=0.35\textwidth]{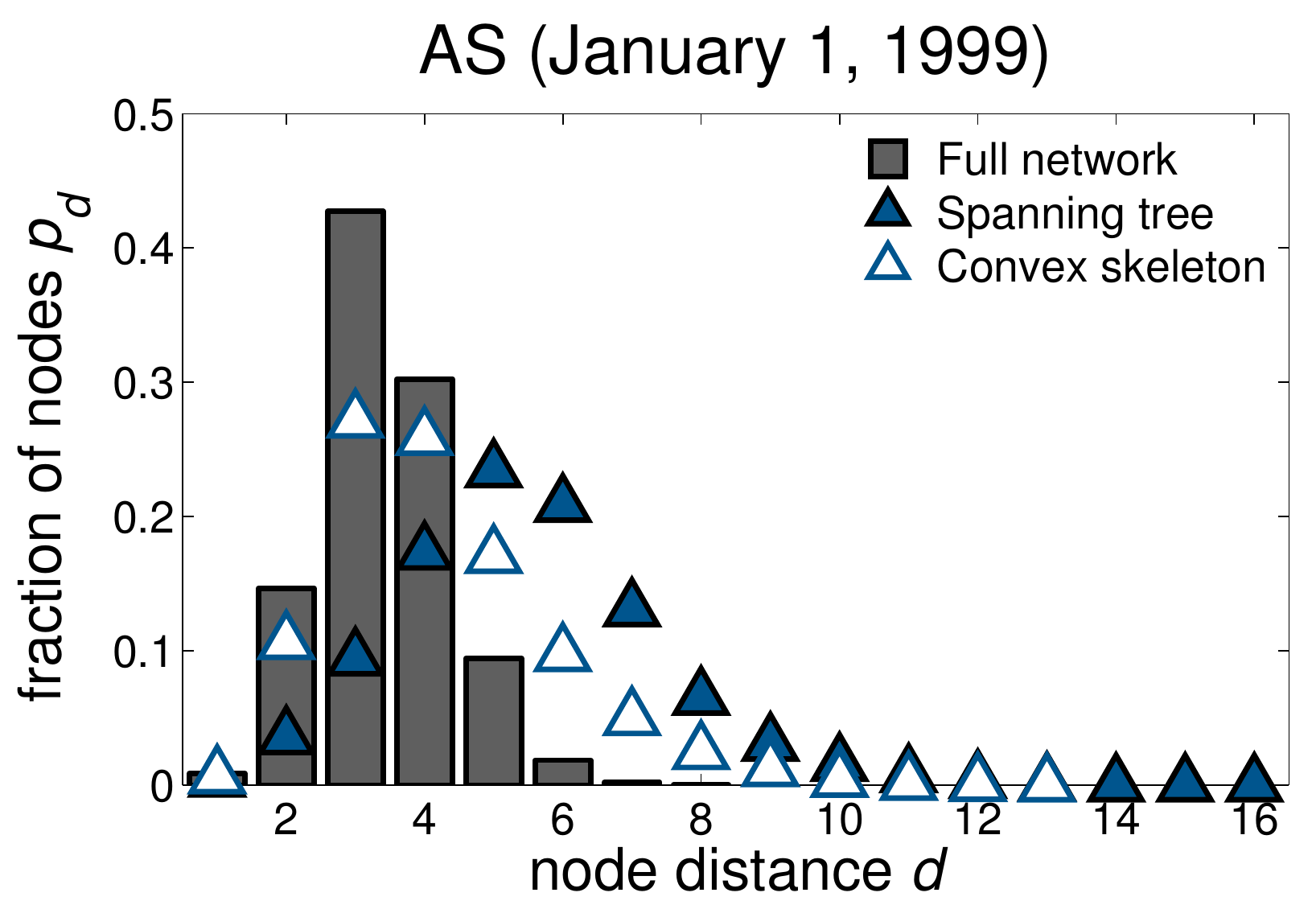}%
	\caption{\label{fig:skeleton}Node and edge distributions of empirical networks, spanning trees and convex skeletons. The plots show the fractions of (\emph{top})~nodes $p_k$ with degree $k$, (\emph{middle right})~edges $p_w$ with weight $w$ and (\emph{bottom})~pairs of nodes $p_d$ at distance $d$ over $25$ independent realisations, whereas the errors bars are not shown. The power-law distributions are maximum likelihood estimations~\cite{CSN09} for the full networks.} 
\end{figure}

Finally, \figref{skeleton} shows different node distributions of the extracted convex skeletons and spanning trees. Notice that the convex skeletons preserve the distributions of node degrees $p_k$ in the full networks (top row), regardless of whether these are heavy-tailed or contain a clear peak. Moreover, the convex skeletons largely retain the distributions of distances between the nodes $p_d$ (bottom row). In contrast to the convex skeletons, the lack of cliques in the spanning trees noticeably increases the distances, while low-degree nodes become overrepresented and high-degree nodes become underrepresented. This is best observed in the case of the network scientists coauthorship network (\diams and \diame). In~\appref{nodes}, we compare the node distributions of convex skeletons and spanning trees also in larger networks, where the conclusions are exactly the same.

%
%

\section{\label{sec:apps}Applications of convex skeletons}

A convex skeleton represents a simple definition of a network backbone~\cite{HLMSW16,CN17} with a very plain structure consisting only of cliques and a tree. Nevertheless, as shown in~\secref{skeleton}, convex skeletons retain important structural properties of empirical networks. In this section, we more thoroughly analyse also node position and community structure of convex skeletons, and compare different backbones extracted from the Slovenian computer scientists coauthorship network.

\subsection{Node position in convex skeletons}

\begin{figure}[t]
	\centering\includegraphics[height=0.3\textwidth,trim={2.5cm 5.75cm 2.25cm 5.75cm}]{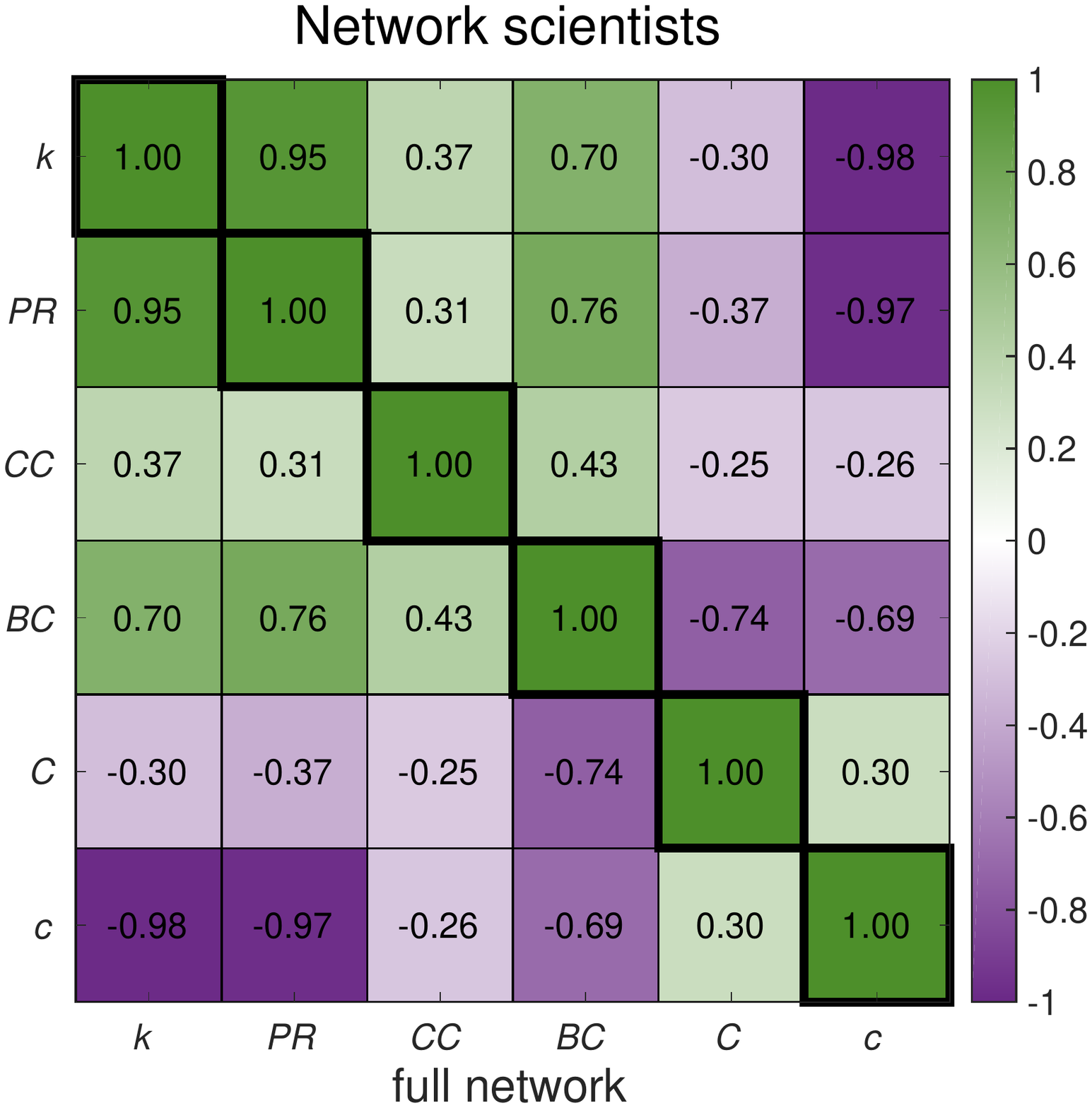}\hskip0.033\textwidth%
	\includegraphics[height=0.3\textwidth,trim={2.5cm 5.75cm 2.25cm 5.75cm}]{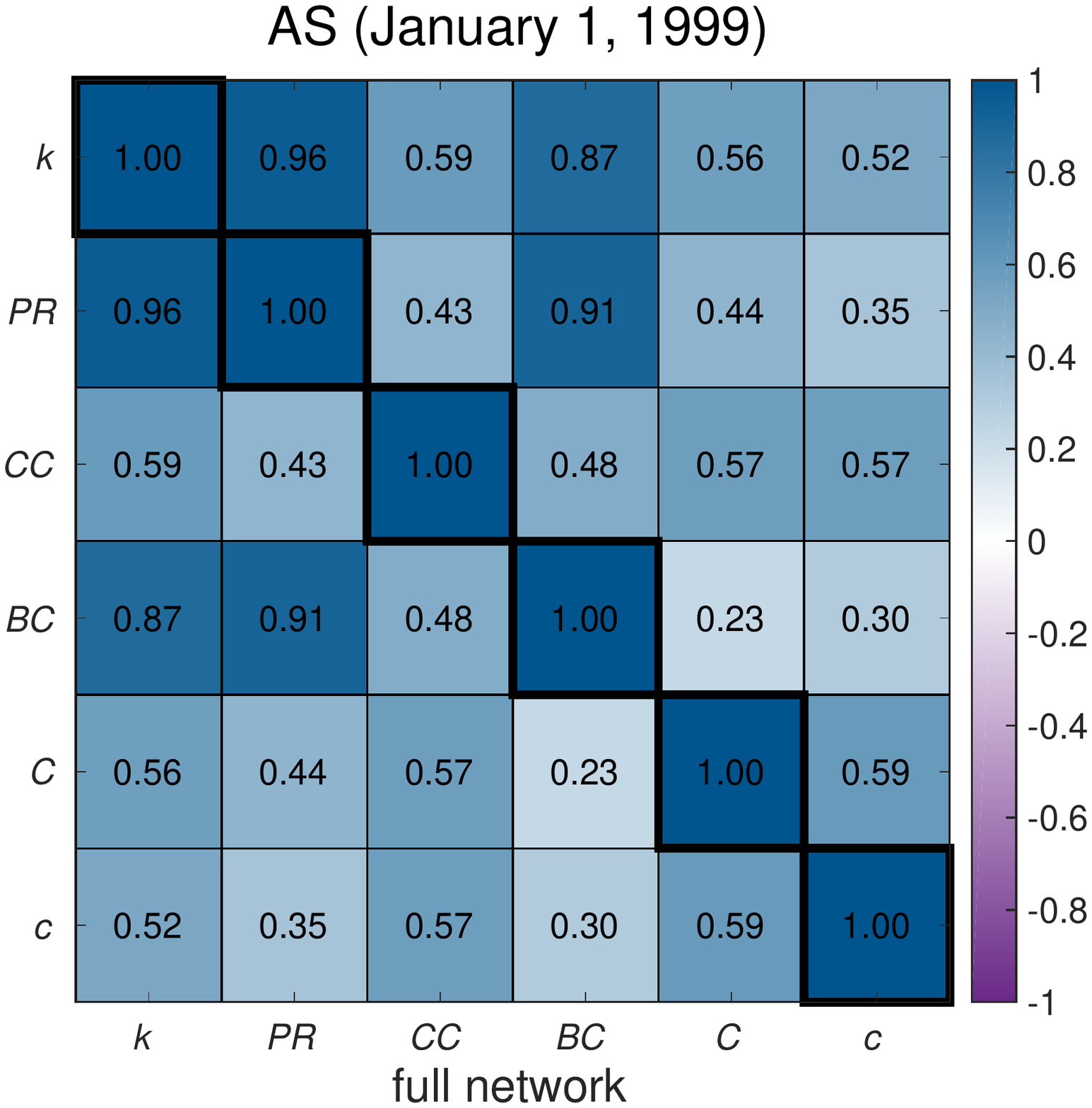}\hskip0.033\textwidth%
	\includegraphics[height=0.3\textwidth,trim={2.5cm 5.75cm 2.25cm 5.75cm}]{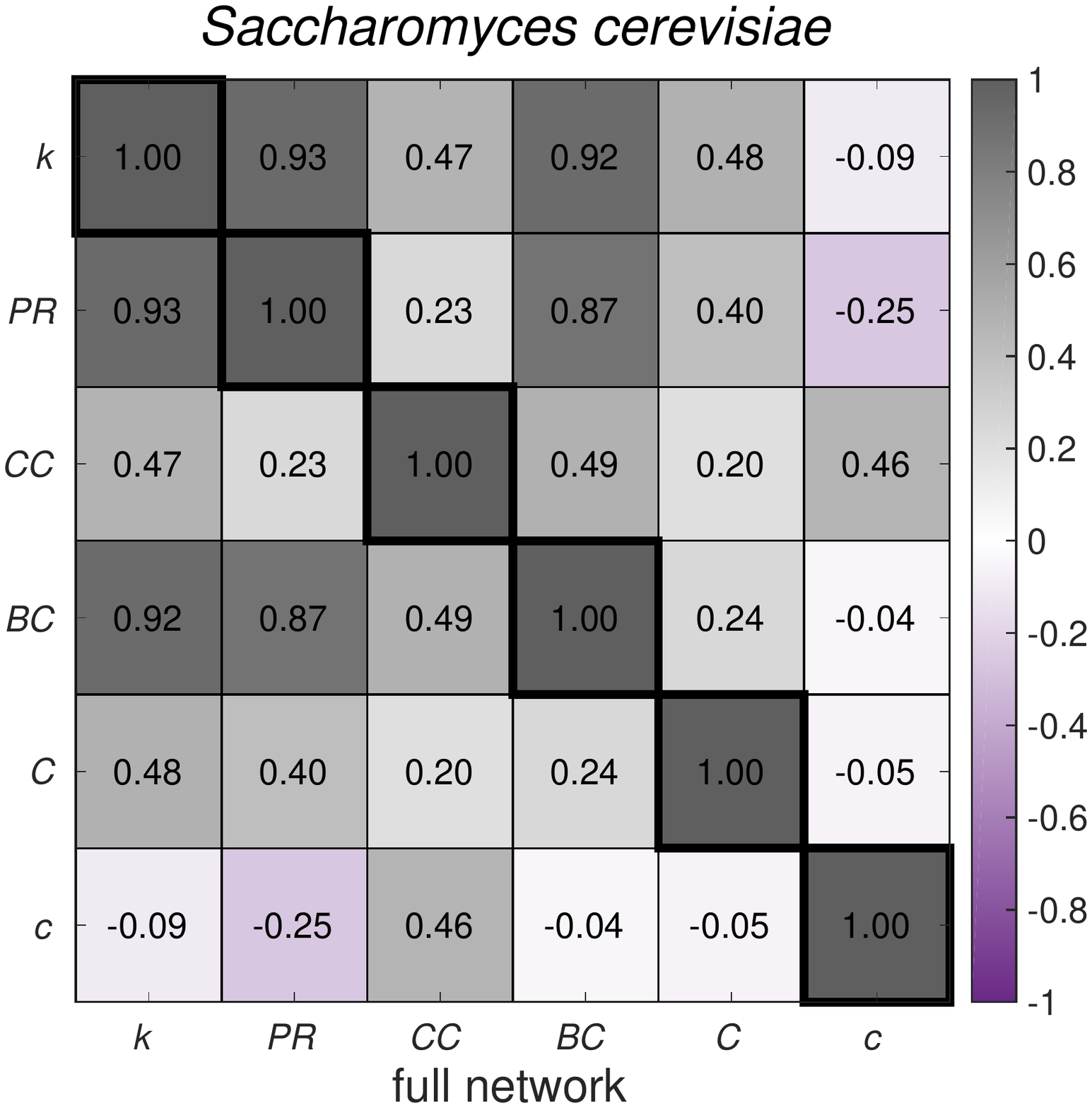}\\\vskip0.04\textwidth%
	\includegraphics[height=0.3\textwidth,trim={3.25cm 5.75cm 2.75cm 5.75cm}]{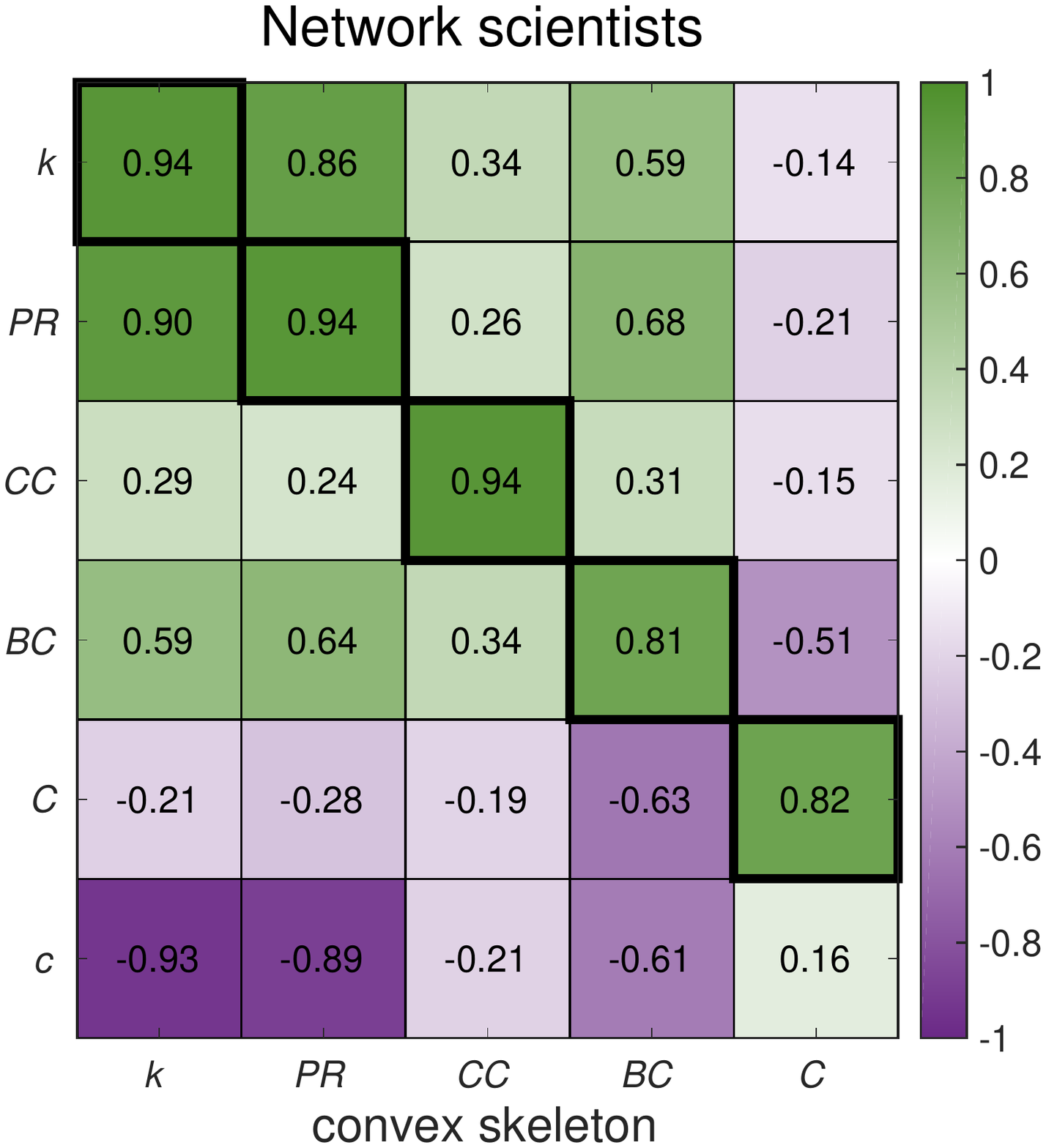}\hskip0.055\textwidth%
	\includegraphics[height=0.3\textwidth,trim={3.25cm 5.75cm 2.75cm 5.75cm}]{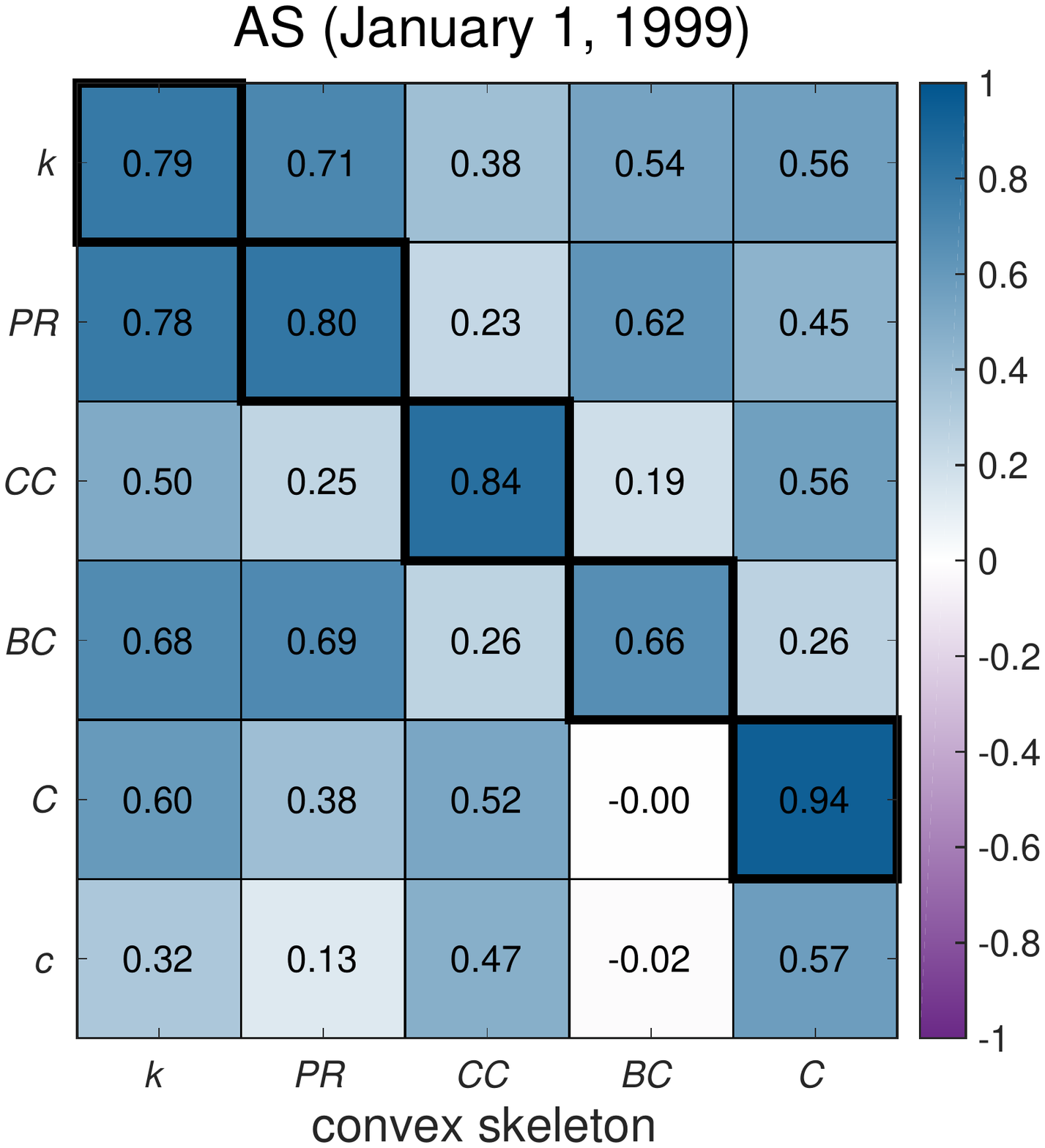}\hskip0.055\textwidth%
	\includegraphics[height=0.3\textwidth,trim={3.25cm 5.75cm 2.75cm 5.75cm}]{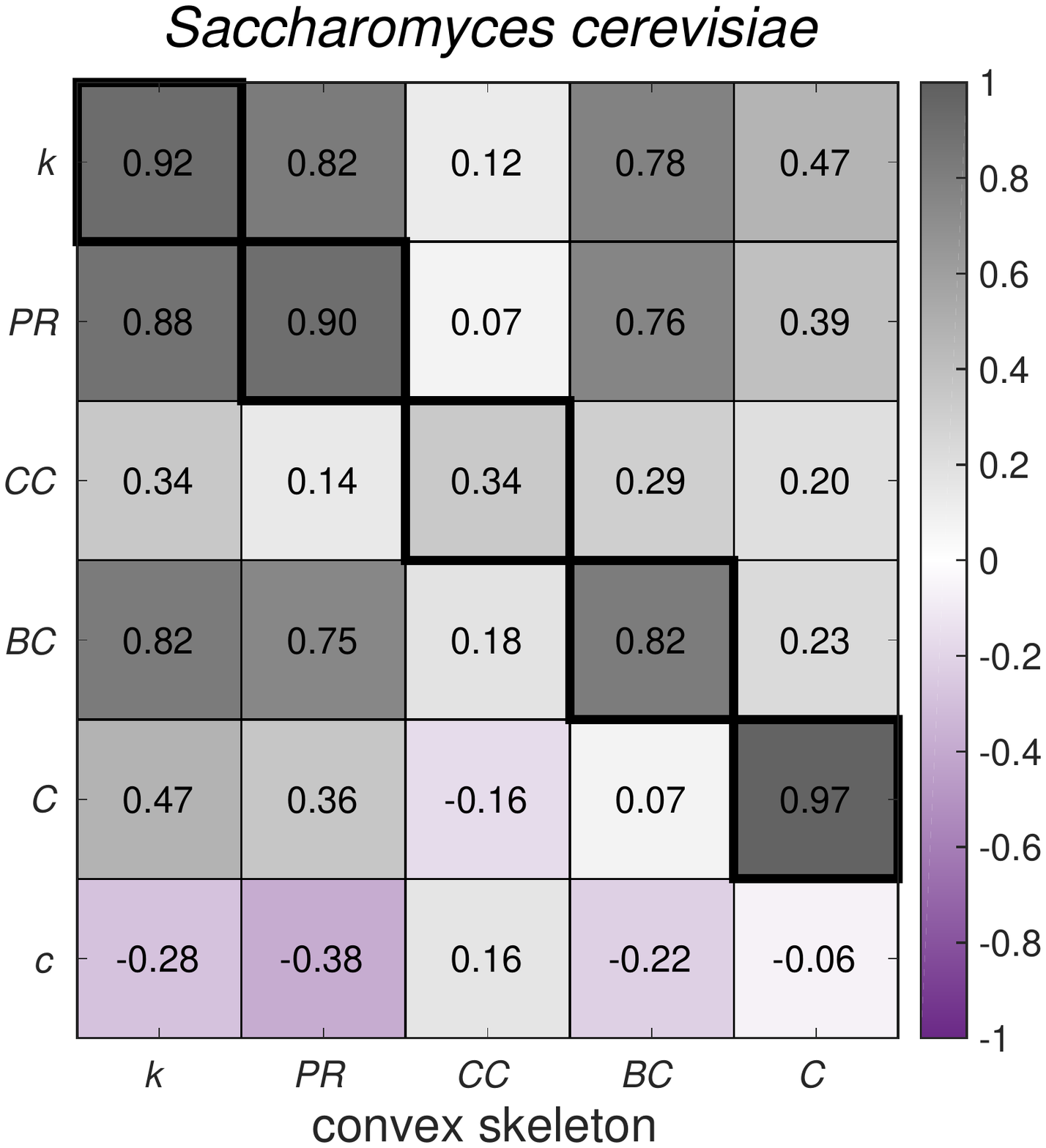}\\\vskip0.04\textwidth%
	\includegraphics[height=0.3\textwidth,trim={3.25cm 5.75cm 2.75cm 5.75cm}]{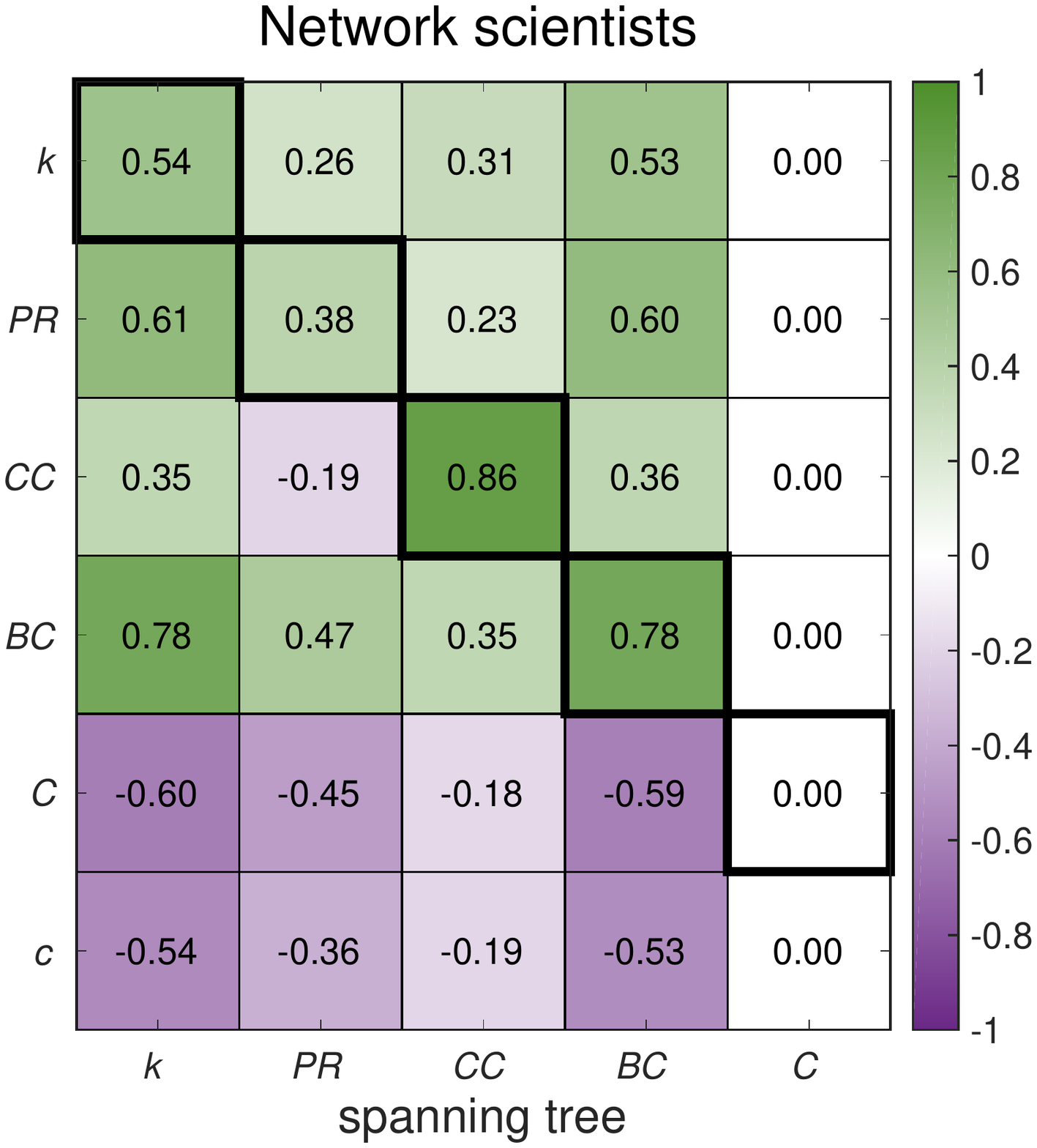}\hskip0.055\textwidth%
	\includegraphics[height=0.3\textwidth,trim={3.25cm 5.75cm 2.75cm 5.75cm}]{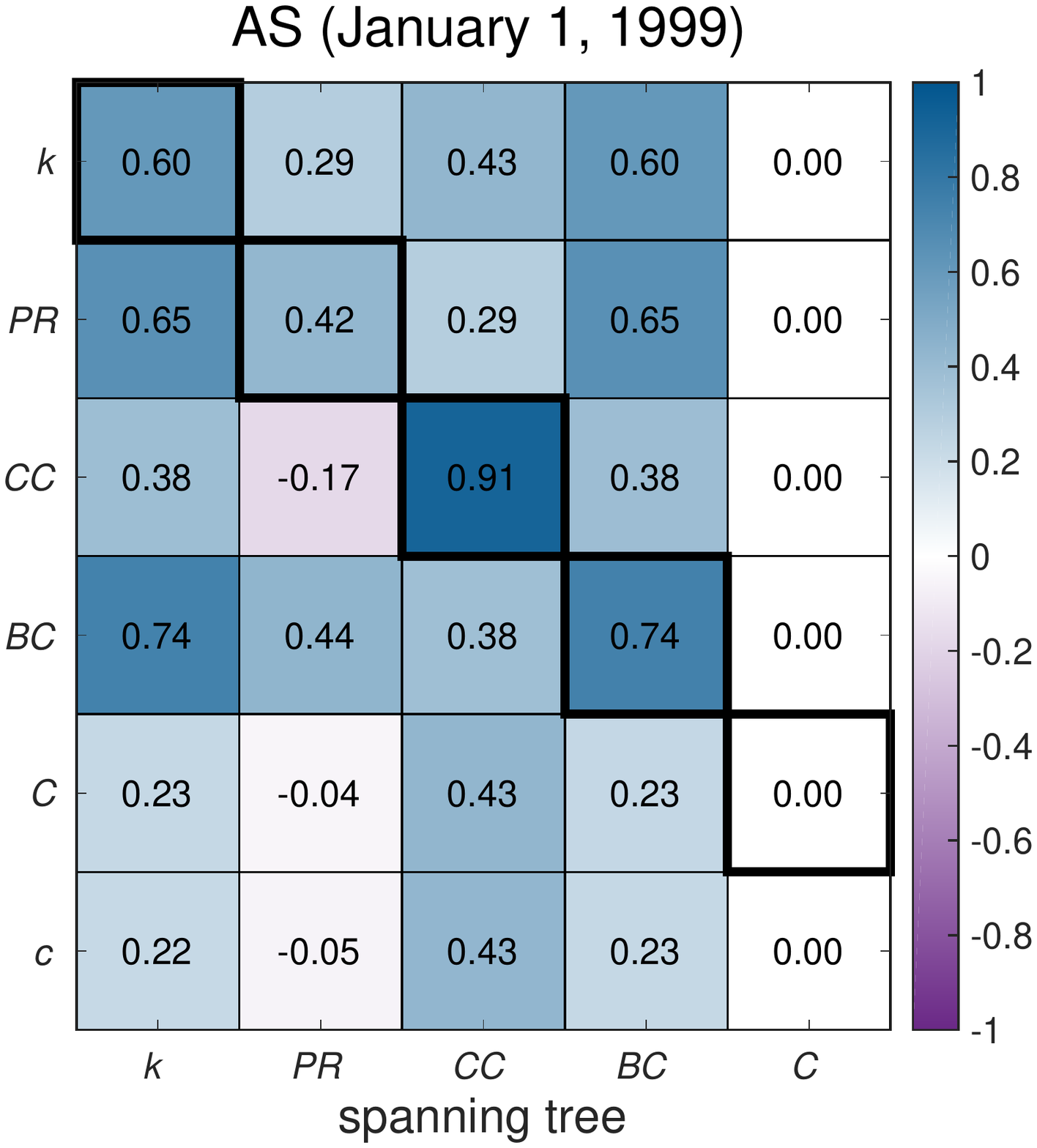}\hskip0.055\textwidth%
	\includegraphics[height=0.3\textwidth,trim={3.25cm 5.75cm 2.75cm 5.75cm}]{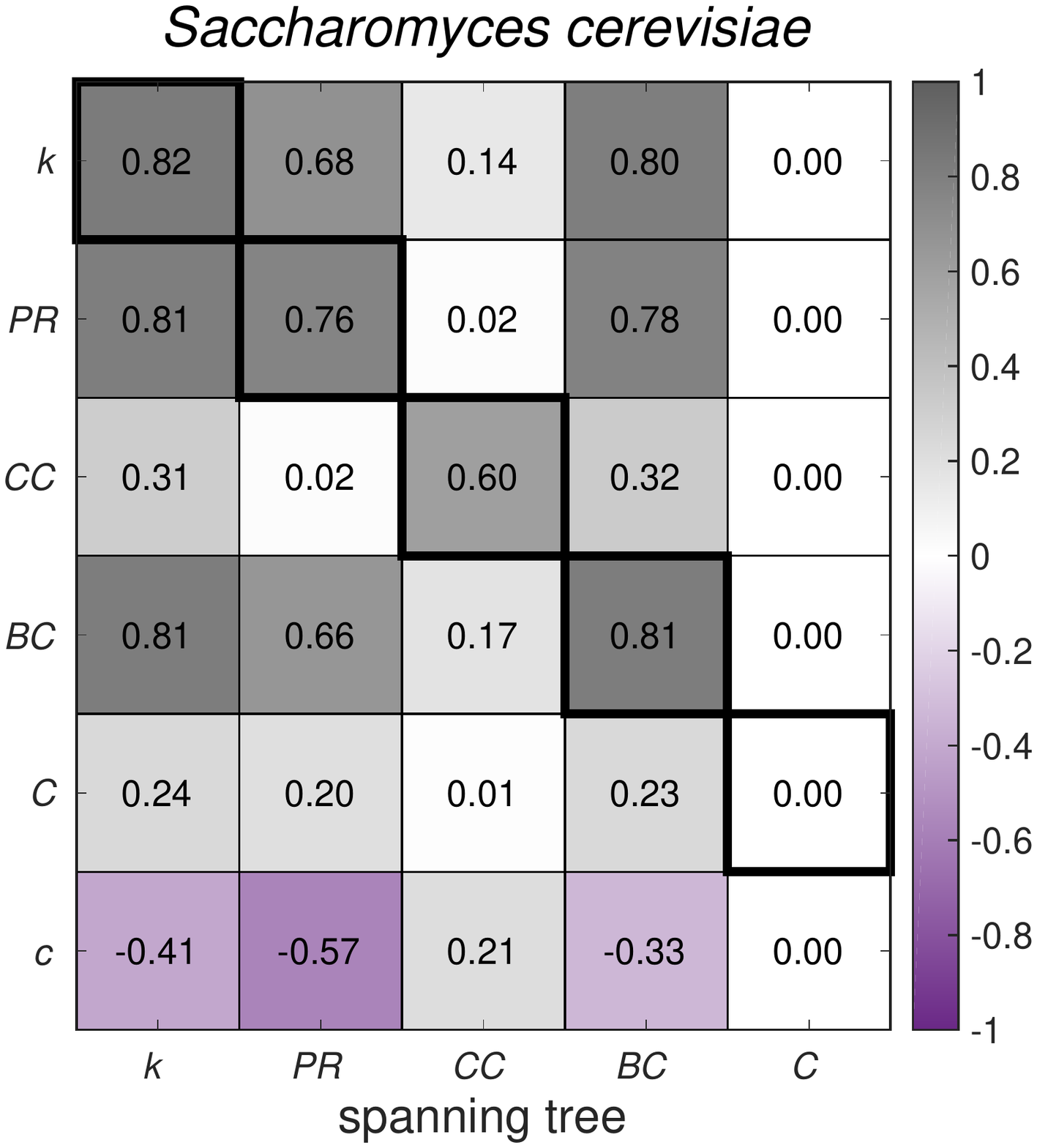}%
	\caption{\label{fig:centrality}Heat maps showing Pearson correlation coefficient between the measures of node position (\emph{top})~in empirical networks, and between networks and (\emph{middle})~convex skeletons or (\emph{bottom})~spanning trees in rows and columns, respectively. The measures include the node degree $k$ and PageRank score {\it PR}, the closeness and betweenness centralities {\it CC} and {\it BC}, and the clustering coefficient $C$ and c-centrality $c$. Note that c-centrality $c$ is only reasonable for full networks and is therefore omitted for convex skeletons and spanning trees. The values are averages over $25$ independent realisations.} 
\end{figure}

\Figref{centrality} shows the Pearson correlation coefficient between the measures of node position in the network scientists coauthorship network, the autonomous systems graph from 1999 and the \scere protein interactions network. The networks are more thoroughly described in~\appref{nets}. The measures include different node centralities and link analysis algorithms. These are node degree and link analysis algorithm PageRank~\cite{BP98} defined as the probability that a random walker with teleports visits a node. Next, we consider geodesics-based measures called node closeness centrality~\cite{Bav50} defined as the average reciprocal distance to other nodes and betweenness centrality~\cite{Fre77} defined as the fraction of geodesics between all nodes that pass through a node. Last, we include also node clustering coefficient~\cite{WS98} already introduced in \secref{skeleton} and c-centrality defined in~\eqref{c}.

The top row of~\figref{centrality} shows that spectral centrality measures node degree and PageRank score are highly correlated, as it is common in undirected networks. Similar holds for betweenness centrality. On the other hand, closeness centrality and clustering coefficient capture different aspects of node position in these networks. Finally, c-centrality is negatively correlated with spectral and betweenness centralities in the coauthorship network, moderately correlated with most measures in the autonomous systems graph, while no strong correlations are observed in the protein interactions network. Hence, c-centrality introduced in this paper is different from standard measures of centrality and link analysis, capturing different aspects of node position.

The middle row of~\figref{centrality} further shows correlations between the node position in empirical networks and convex skeletons. Notice that all considered aspects of node position are retained in convex skeletons with values of the correlation coefficient between $0.8$ and $0.95$ in most cases. In contrast, the center of the protein interactions network shifts in the convex skeletons resulting in weak correlation $0.34$ of closeness centrality. In the bottom row of~\figref{centrality}, we also compare the node position in networks to spanning trees, which do not retain the clustering. Consequently, the correlation of clustering coefficient is undefined shown as empty cells in the heat maps. Similar results are obtained with other network backboning techniques~\cite{Fre77,GTB12,CN17}.

\subsection{Community structure of convex skeletons}

\begin{table}[t]
	\caption{\label{tbl:community}Comparison between community structure of empirical networks (N), and spanning trees (ST) and convex skeletons (CS). The structure is revealed with the map equation algorithm, and compared through the number of communities, normalised mutual information {\it NMI} (higher is better) and normalised variation of information {\it NVI} (lower is better). The values are averages over $25$ independent realisations.}
	\begin{tabular}{llrrrrcccc}
		\multirow{2}{*}{Class} & \multirow{2}{*}{Network} & \multicolumn{3}{c}{\# communities} & \multicolumn{2}{c}{{\it NMI}} & \multicolumn{2}{c}{{\it NVI}} \\
		& & \ncsst & \csst & \csst \\\hline
		\multirow{3}{*}{Collaboration} & \jazz & $6.0$ & $26.6$ & $25.0$ & $0.49$ & $0.36$ & $0.57$ & $0.75$ \\
		& \netsci & $37.3$ & $40.5$ & $50.0$ & $0.96$ & $0.87$ & $0.06$ & $0.22$ \\
		& \cmpsci & $24.0$ & $28.0$ & $31.3$ & $0.88$ & $0.81$ & $0.19$ & $0.33$ \\\hline
		\multirow{3}{*}{\shortstack[l]{Protein\\ interactions}} & \plasm & $138.0$ & $132.8$ & $140.2$ & $0.83$ & $0.77$ & $0.33$ & $0.45$ \\
		& \scere & $162.7$ & $164.1$ & $175.2$ & $0.93$ & $0.91$ & $0.13$ & $0.18$ \\
		& \celeg & $286.7$ & $321.3$ & $348.0$ & $0.86$ & $0.82$ & $0.25$ & $0.32$ \\\hline
		\multirow{3}{*}{\shortstack[l]{Autonomous\\ systems}} & \oreg{1998} & $230.4$ & $285.7$ & $285.3$ & $0.86$ & $0.75$ & $0.25$ & $0.41$ \\
		& \oreg{1999} & $70.2$ & $74.9$ & $80.0$ & $0.76$ & $0.64$ & $0.40$ & $0.61$ \\
		& \oreg{2000} & $271.3$ & $358.8$ & $338.9$ & $0.84$ & $0.74$ & $0.28$ & $0.46$ \\
	\end{tabular} 
\end{table}

We continue by comparing the community structure of empirical networks, and spanning trees and convex skeletons. The networks include different social collaboration and protein interactions networks, and autonomous systems graphs described in~\appref{nets}. The community structure is revealed with the map equation algorithm~\cite{RB08} that identifies communities by compressing the trajectory of a random walker on a network. Note that the algorithm highlights just one specific facet of community structure~\cite{SDRL17}, which has however proved useful in the past~\cite{LF09a,RB11a,PR17}.

\Tblref{community} first shows the number of communities revealed in networks and backbones. These are comparable in all cases but one. In the network of collaborations between jazz musicians, the number of communities in the backbones is considerably higher than in the network, which is due to a  high density of the network and much lower density of the backbones. Next, \tblref{community} shows normalised mutual information~\cite{DDDA05} between the community structures of networks and backbones. In the case of independent structures, the measure equals zero, and is equal to one, when the structures are identical. The results show that the communities in networks are largely retained in the convex skeletons with values between $0.85$ and $0.95$ in most cases. Differently from convex skeletons, the values are consistently lower in the spanning trees. \Tblref{community} also shows the values of normalised variation of information~\cite{Mei07}, which is a measure of distance in the space of partitions, thus lower is better. The values further confirm the observations from above.

\begin{figure}[t]
	\centering\includegraphics[width=0.311\textwidth]{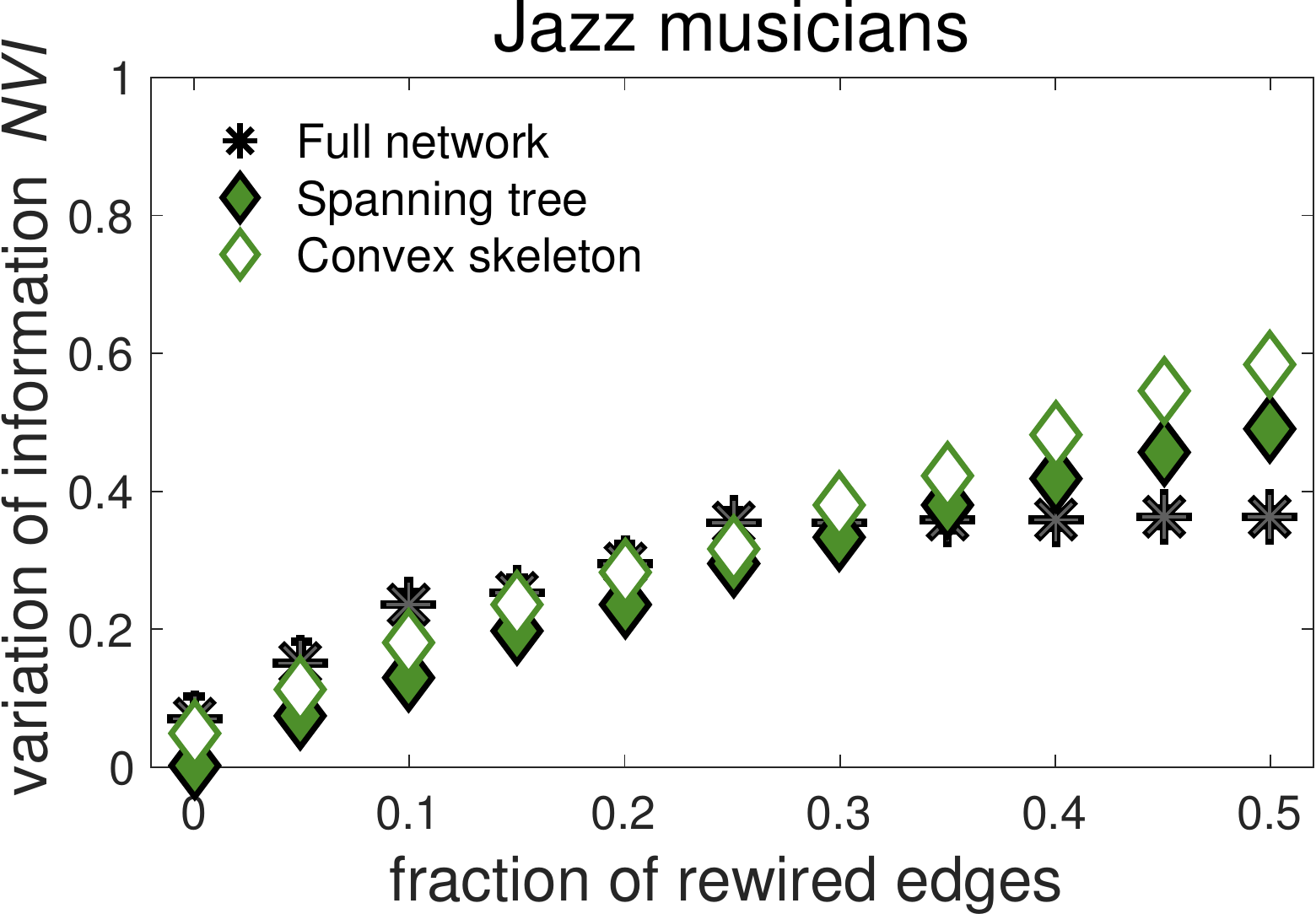}\hskip0.033\textwidth%
	\includegraphics[width=0.311\textwidth]{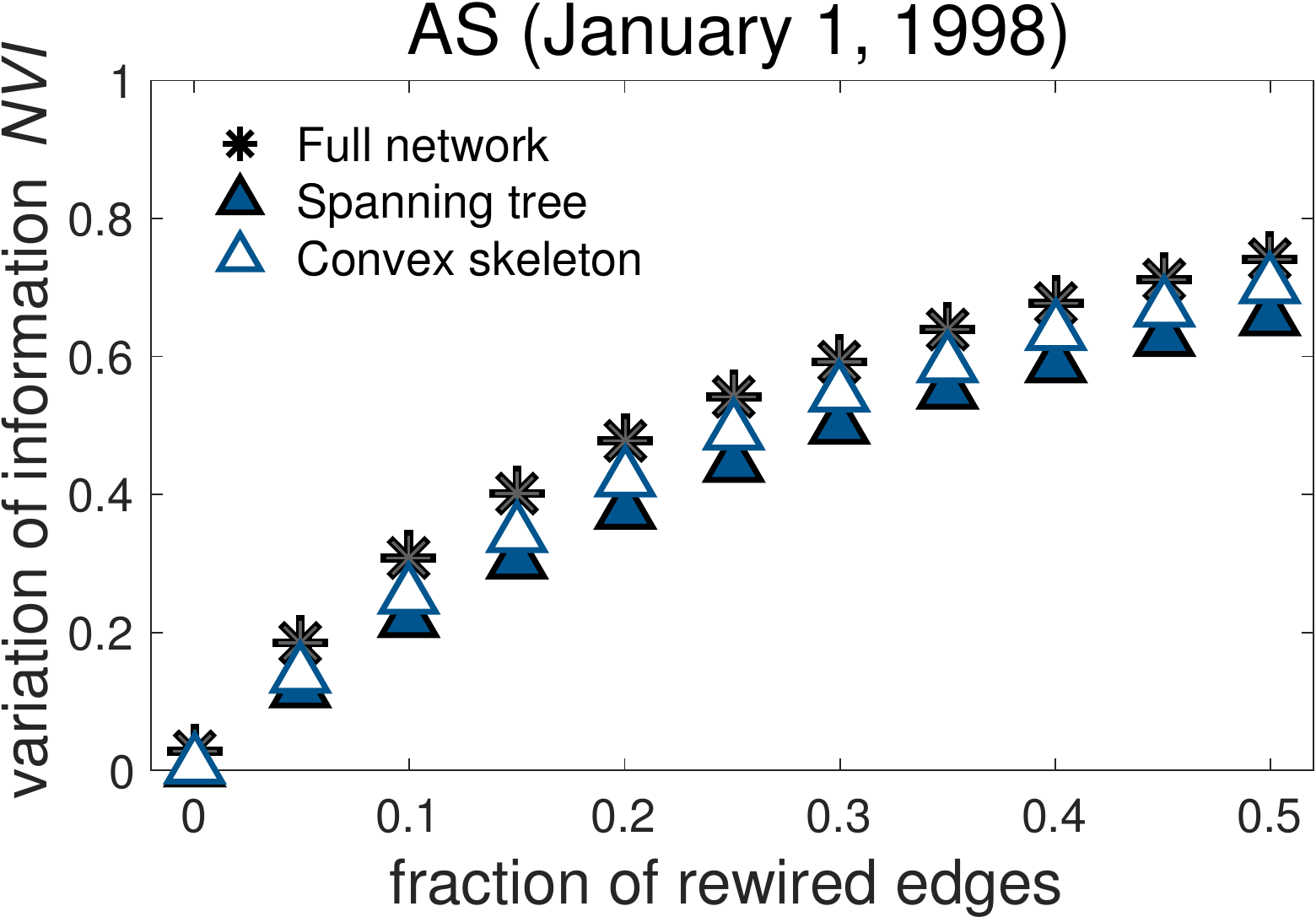}\hskip0.033\textwidth%
	\includegraphics[width=0.311\textwidth]{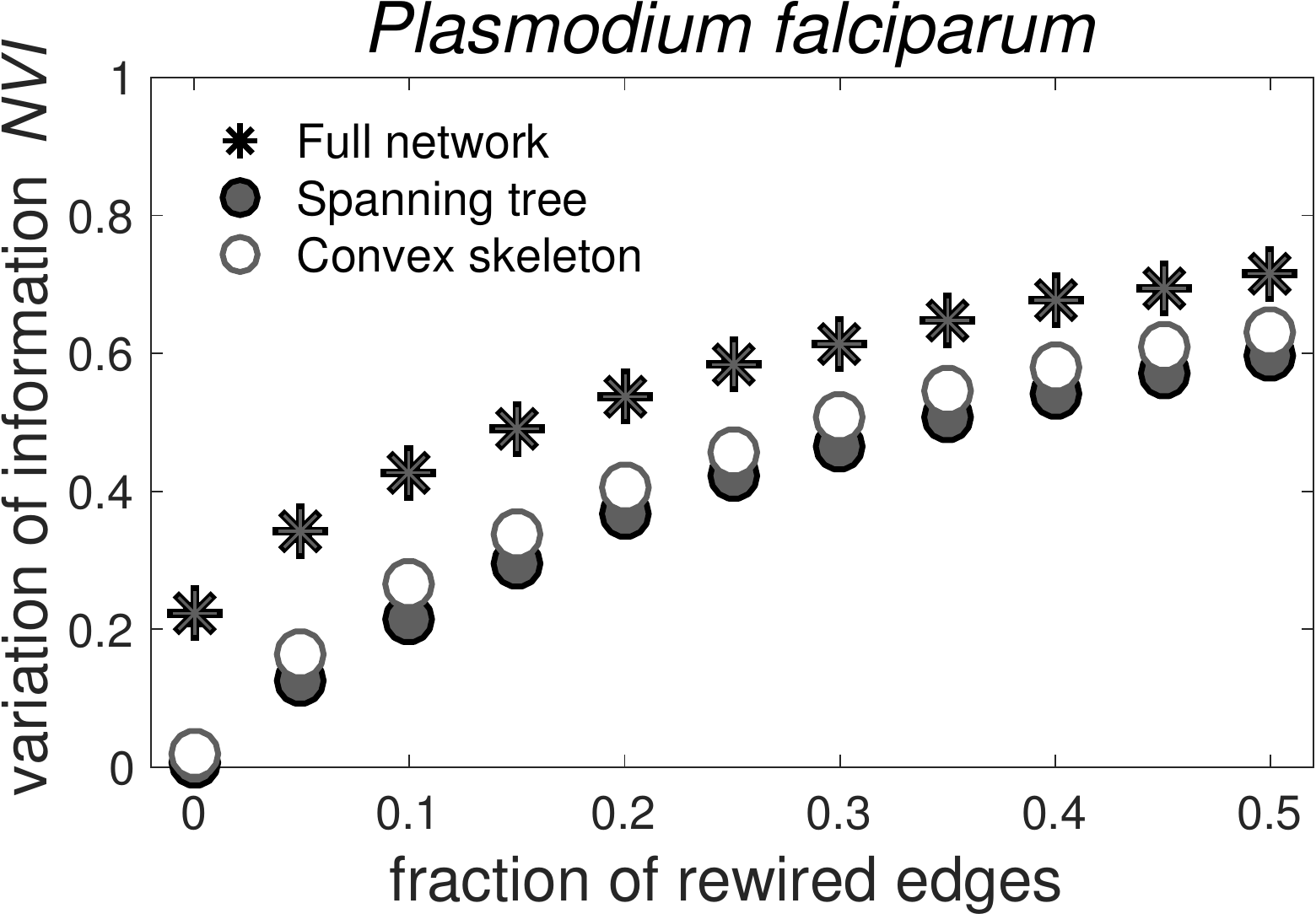}%
	\caption{\label{fig:community}Robustness of community structure of empirical networks, spanning trees and convex skeletons revealed with the map equation algorithm. The plots show the distances between the community structures measured by normalised variation of information {\it NVI} for different fractions of rewired edges. The symbols are averages over $25$ independent realisations, whereas the errors bars are not visible.} 
\end{figure}

\Figref{community} demonstrates also the robustness of community structure in selected networks and backbones. The plots show the distance between community structures under degree-preserving randomisation~\cite{MS02} measured by normalised variation of information. Notice that both backbones strengthen the community structure, making it more robust to random perturbations of the edges. This is most notably observed in the protein interactions network. In summary, convex skeletons not only preserve the community structure of empirical networks, but at the same time make the structure more pronounced. In the remaining of the section, we further investigate this in the case of a coauthorship network.

\subsection{Network backbones of coauthorship network}

Convex skeleton seem to provide a particularly reasonable abstraction of different social collaboration networks such as coauthorship networks~\cite{New01a}, since these are unions of cliques by construction. In the following, we analyse convex skeletons extracted from the Slovenian computer scientists coauthorship network~\cite{BSB12} and compare them with other network backboning techniques.

The coauthorship network was collected from the Slovenian national research database SICRIS~\cite{sicris} in 2010. We here analyse only the largest connected component of the network representing coauthorships between $239$ computer scientists, which is about a third of all registered computer scientists at that time. Since each scientists in the SICRIS database is represented by a unique identifier, no name disambiguation was needed. Note, however, that the main purpose of the SICRIS database is not the archiving of scientific contributions, but the evaluation of the performance of scientists, which may reflect certain policy needs~\cite{Rod16}. Further details about the network are given in~\appref{nets}.

\Tblref{coauthor} shows the standard statistics of the network, and the extracted spanning trees and convex skeletons. The results are consistent with~\secref{skeleton}. While both techniques produce a high-convexity backbone with $Xs\geq 0.95$ and retain the connectivity of the network $s\approx 1$, only the convex skeletons preserve the average node clustering coefficient $\avg{C}=0.54$ and to a large extent also the distances between the nodes $\avg{\ell}=7.06$. Moreover, the convex skeletons retain the node degree distribution $p_k$ of the network (\diame in the middle row~of~\figref{skeleton}).

\begin{table}[t]
	\caption{\label{tbl:coauthor}Summary statistics of backbones of the computer scientists coauthorship network. These are the average node degree $\avg{k}$ and clustering coefficient $\avg{C}$, the average distance between the nodes $\avg{\ell}$, modularity $Q$ of the classification of authors into fields, corrected convexity $Xs$ and the fraction of nodes in the largest connected component $s$. The values are averages over $25$ independent~realisations.}
	\begin{tabular}{lcrcccc}
		Backbone & $\avg{k}$ & \mc{$\avg{\ell}$} & $\avg{C}$ & $Q$ & $Xs$ & $s$ \\\hline
		Full network & $4.75$ & $4.58$ & $0.48$ & $0.33$ & $0.64$ & $1.00$ \\
		Spanning tree & $1.99$ & $10.11$ & $0.00$ & $0.32$ & $1.00$ & $1.00$ \\
		\multirow{2}{*}{Edge betweenness} & $3.25$ & $4.63$ & $0.18$ & $0.29$ & $0.69$ & $1.00$ \\
		& $3.25$ & $2.44$ & $0.40$ & $0.36$ & $0.35$ & $0.64$ \\
		Salience skeleton & $3.10$ & $4.78$ & $0.05$ & $0.30$ & $0.57$ & $1.00$ \\
		Convex skeleton & $3.25$ & $7.06$ & $0.54$ & $0.38$ & $0.95$ & $0.99$ \\
	\end{tabular}
\end{table}

\Tblref{coauthor} further shows the statistics of two alternative network backboning techniques based on edge betweenness~\cite{Fre77} and salience~\cite{GTB12}. The betweenness of an edge is defined as the fraction of shortest paths between all nodes that traverse the edge, while the salience of an edge is defined as the fraction of shortest path trees originating from each node that include the edge. Consistently with a convex skeleton, both techniques define a network backbone based on the shortest paths between the nodes. In contrast, the techniques are based on the number of shortest paths and not their inclusion as in a convex skeleton. The salience of the edges was computed using the implementation provided in~\crefs{CN17,backbone}, while we use our own implementation for edge betweenness~\cite{Bra01}.

The distribution of edge salience is bimodal~\cite{GTB12}, which gives a natural definition of a high-salience skeleton. On the other hand, the distribution of edge betweenness is heavy-tailed in the concerned coauthorship network. In~\tblref{coauthor}, we therefore report the statistics of a high-betweenness backbone with the same number of edges as in the convex skeletons (first row) and, for comparison, also a low-betweenness backbone (second row).

Consider first the high-salience skeleton and the high-betweenness backbone. Both techniques retain the connectivity of the network $s=1$ and short distances between the nodes $\avg{\ell}\approx 4.71$, while corrected convexity remains comparable with the full network $Xs\approx 0.63$. Yet, the average node clustering coefficient drops to $\avg{C}\approx 0.12$, since the backbones do not preserve the cliques in the network. This is because the edges embedded in the cliques are not bridges between different parts of the network that would support many shortest paths. These edges are in fact included in the low-betweenness backbone with $\avg{C}=0.40$, but which disconnects the network as $s=0.64$. In summary, out of all the network backbones considered, only the convex skeletons retain the important structural properties of this coauthorship network. Besides, the convex skeletons preserve also the power-law distribution of the weights of the coauthorship ties $p_w$, where the weight $w$ of a tie is defined as the number of coauthored papers (\squre in the middle row~of~\figref{skeleton}).

\begin{figure}[t]
	\centering\includegraphics[width=0.7\textwidth,trim={2.5cm 5cm 2.5cm 5cm},clip]{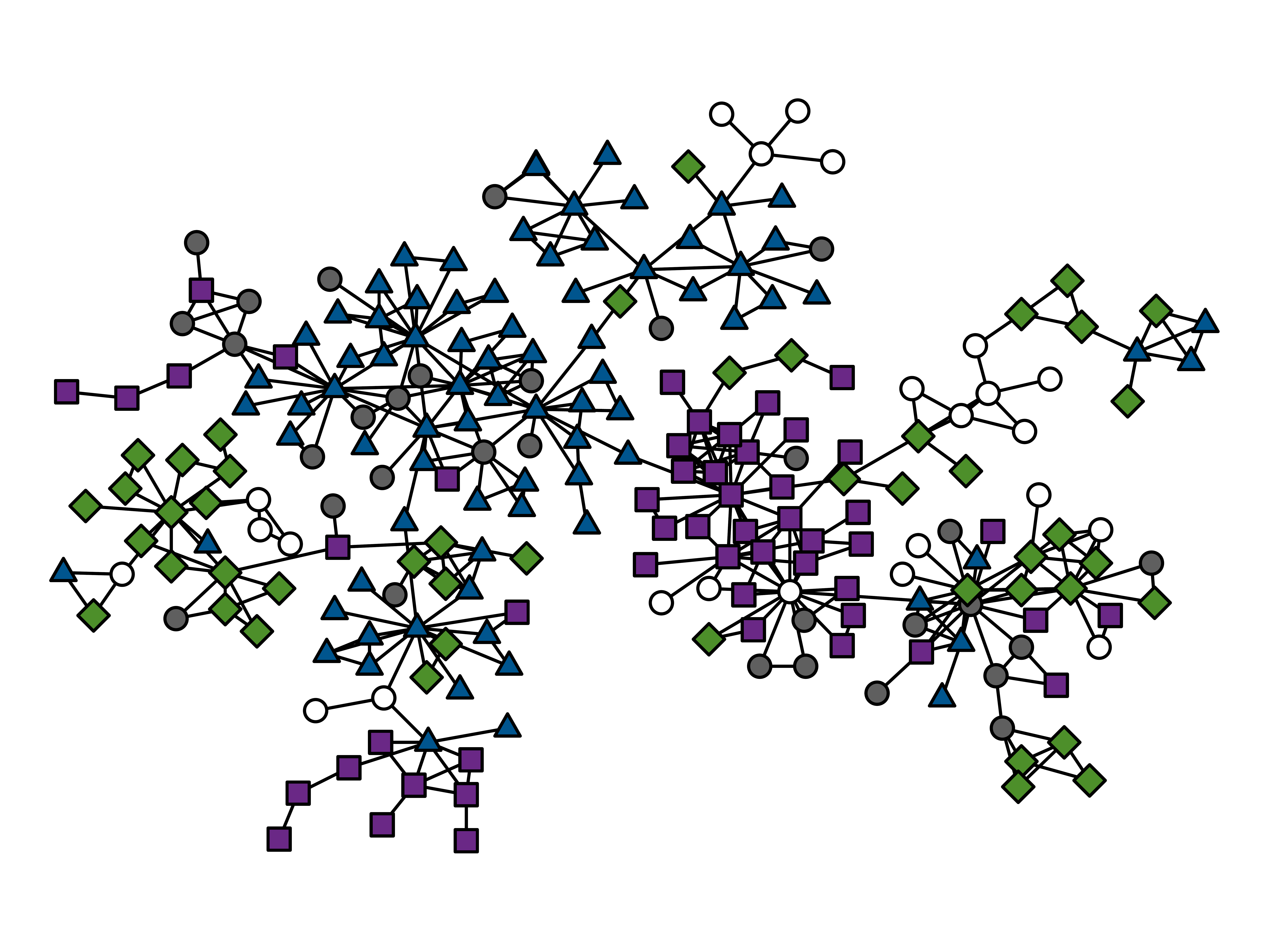}\\\vskip0.025\textwidth%
	\includegraphics[height=0.225\textwidth,trim={0 0 6.5cm 0},clip]{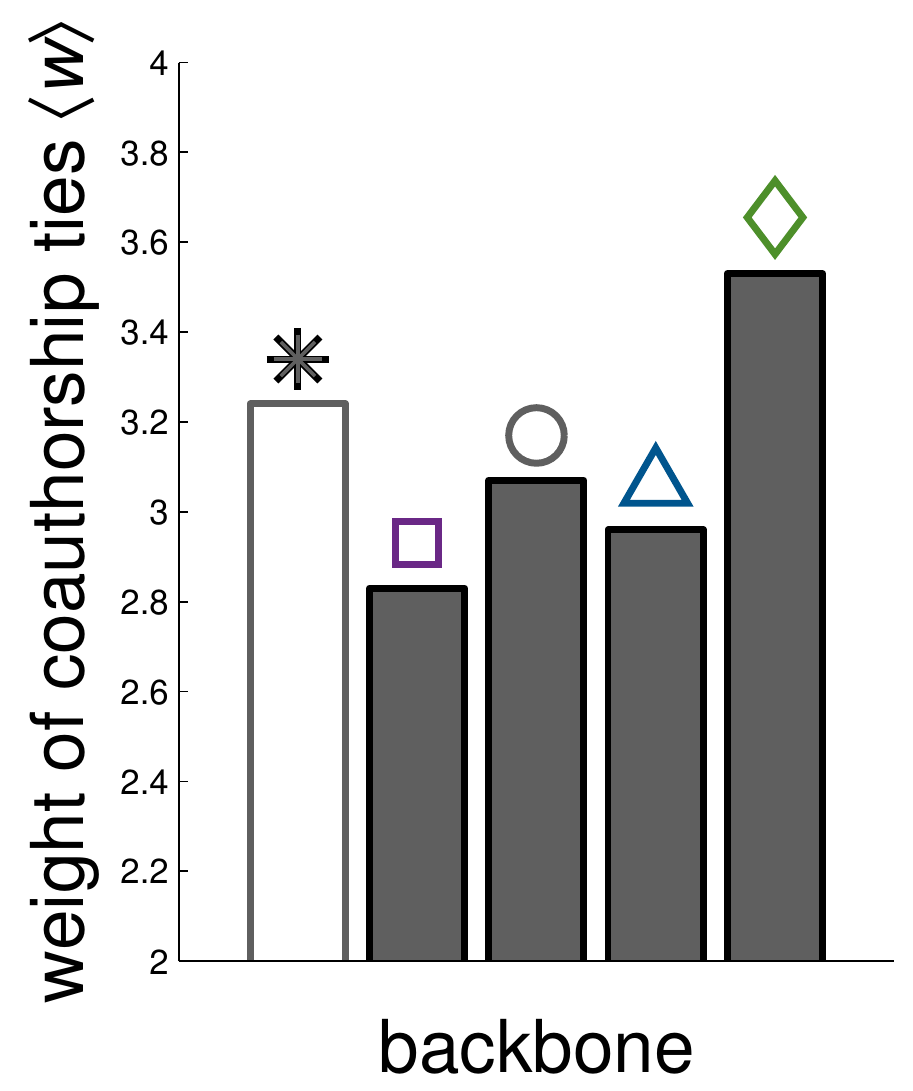}\hskip0.025\textwidth%
	\includegraphics[height=0.225\textwidth,trim={0 0 6.5cm 0},clip]{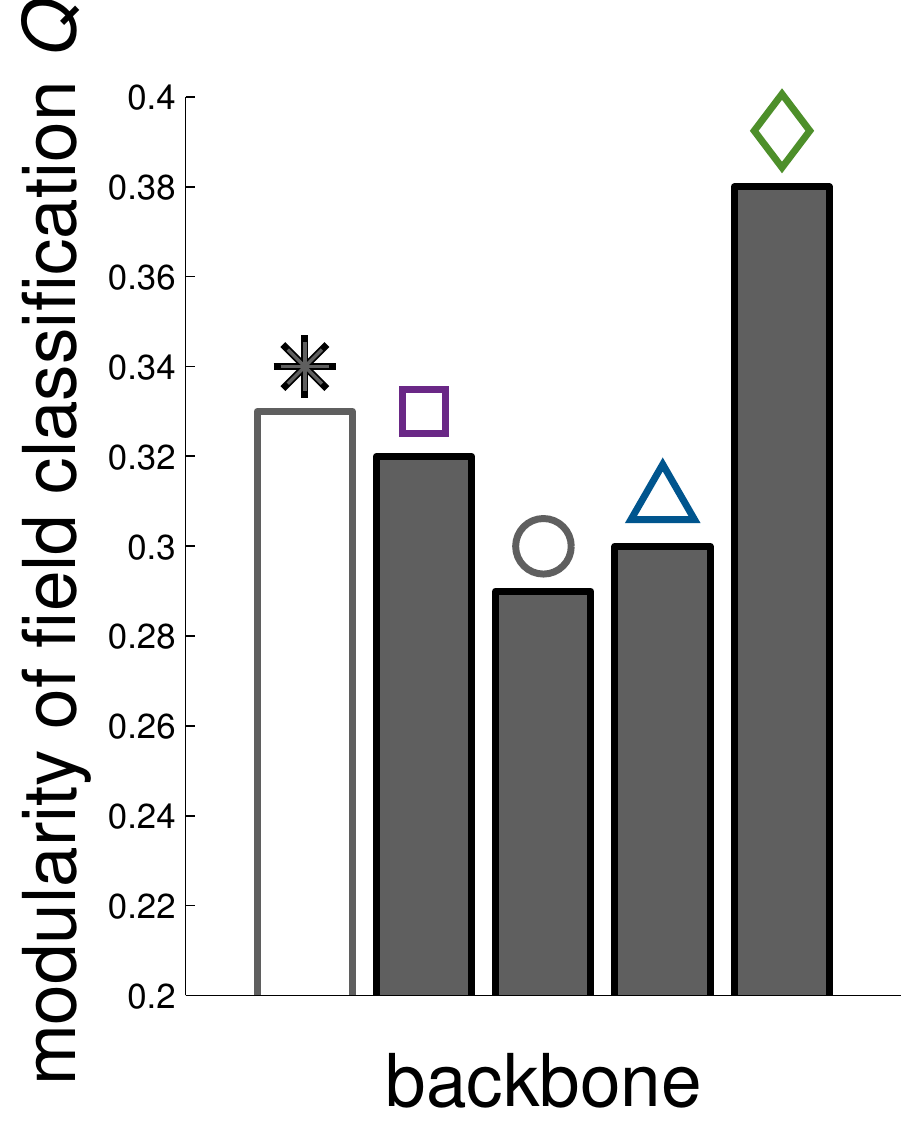}\hskip0.025\textwidth%
	\includegraphics[height=0.225\textwidth]{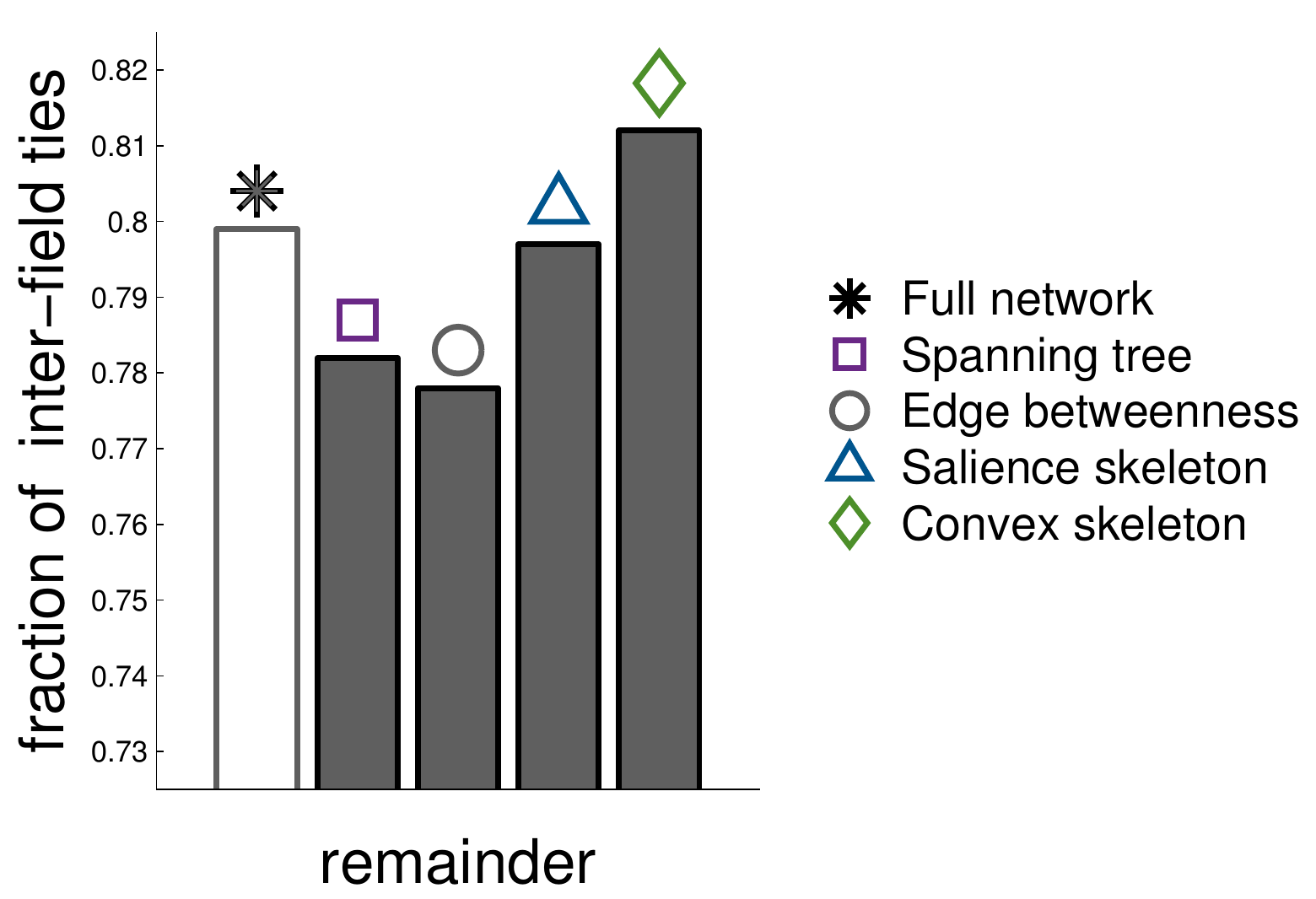}%
	\caption{\label{fig:compsci}(\emph{top})~A convex skeleton of the computer scientists coauthorship network. The node symbols represent primary research fields of the authors: computer theory (\diams), information systems (\squrs), intelligent systems (\trians), programming technologies (\ellpe) and other (\ellps). (\emph{bottom})~Comparison of network backbones showing the average weight $\avg{w}$ of the coauthorship ties in a backbone and modularity $Q$ of the field classification, and the fraction of inter-field ties in the remaining network. The histograms are averages over $25$ independent realisations.} 
\end{figure}

A realisation of a convex skeleton is shown in the top of~\figref{compsci}, where node symbols represent the primary research fields of the authors. One can visually observe the underlying tree structure of the convex skeleton and also the cliques formed by the authors from the same research field. Furthermore, the bottom of~\figref{compsci} compares the properties of different backbones that one might find desirable for an abstraction of a coauthorship network. These are the average weight of the coauthorship ties $\avg{w}$ in the backbones, modularity $Q$~\cite{GN02} of the field classification that compares the number of ties between the authors from the same field with the expected number of such ties and the average fraction of inter-field ties in the remaining network after removal of the backbones. In all cases, these properties are enhanced in a convex skeleton (\diame) compared to the full network, differently from a spanning tree (\squre), high-betweenness backbone (\ellpe) and also high-salience skeleton (\triane). Preliminary results on coauthorship networks from the SICRIS database for other scientific disciplines are consistent with above, while detailed results will be reported elsewhere.

%
%

\section{\label{sec:conc}Conclusions}

In this paper, we have introduced a network backboning technique called a convex skeleton. A convex skeleton is a collection of cliques glued together in a tree and as such represents a generalisation of a network spanning tree. We have extracted convex skeletons from empirical networks of various type and origin, and shown that these retain the degree distribution, clustering, connectivity, distances, node position and also community structure, while at the same time make the shortest paths in a network largely unique. This is in contrast with a spanning tree or high-betweenness backbone~\cite{Fre77} and high-salience skeleton~\cite{GTB12}. A convex skeleton seems to provide a particularly reasonable abstraction of social collaboration networks such as coauthorship networks, with applications in network analysis, modelling, visualisation, navigation and possibly also elsewhere. Still, convex skeletons are not the most reasonable abstraction of networks that are a priori non-convex like food webs or random graphs.

A convex skeleton can be seen as a particular case of a network sampling or sparsification technique~\cite{LF06,HLMSW16,BSB17}, but with diverging goals. At the one hand, sparsification techniques try to simplify the network by removing redundant or spurious edges~\cite{GS09b,CN17}, while at the same time keep its structural properties intact. Therefore, one would ideally like to remove as many edges as possible. On the other hand, a convex skeleton tries to retain as many edges as possible, since the simplicity is implied by its plain structure. It should, however, be mentioned that the approach for extracting convex skeletons adopted here is just one example, and likely much more efficient and effective techniques exist. The important fact is that one does not have to estimate convexity to extract a convex skeleton. Consequently, the most expensive part of the approach is ensuring that convex skeletons remain connected, which could be ensured in lesser time by using recent techniques for detecting bridges in networks~\cite{WTL18}. Besides, we were here interested only in revealing a convex skeleton by removal of edges, while adding edges in order to obtain a fully convex network might be more suitable for some applications.

As stated above, the concept of convexity is not actually needed in order to extract a convex skeleton. Then, why bother the reader with the definition of convexity in the first place? The reason is that convexity allows one to measure how far the resulting structure is from a tree of cliques, which would be hard to estimate otherwise. Finally, it is reasonable to expect that trees of cliques have been studied in networks context before, either in graph theory or network science literature. Nevertheless, we are convinced that our definition based on convexity is novel and unique. We are eager to see whether the ideas developed in this paper will prove useful for better understanding of network structure and stimulate new applications of network theory.

%
%

\vskip1pc

\dataccess{The networks used in this work are freely available online within the SNAP datasets at \url{http://snap.stanford.edu/data}, the Pajek datasets at \url{http://vlado.fmf.uni-lj.si/pub/networks/data}, the KONECT repository at \url{http://konect.uni-koblenz.de} and the BioGRID repository at \url{https://thebiogrid.org}. The codes used for network convexity are available at \url{https://github.com/t4c1/Graph-Convexity}, the codes for convex skeletons at \url{https://github.com/t4c1/Convex-Skeleton}, the codes for network backboning at \url{http://www.michelecoscia.com/?page_id=287} and the codes for community detection at \url{http://www.mapequation.org/code.html}.}

\competing{The authors have no conflicting interests to declare.}

\funding{This work has been supported in part by the Slovenian Research Agency under the program P2-0359 and by the European Union COST Action number CA15109.}

\ack{The authors thank Tilen Marc, Vincent Traag, Ludo Waltman, Zoran Levanji\'{c}, Matja\v{z} Perc, Dalibor Fiala and two anonymous reviewers for helpful comments and suggestions, Tadej Ciglari\v{c} for implementations of network convexity and convex skeletons, Michele Coscia for implementations of network backbones and Martin Rosvall for implementations of community detection algorithms. The authors also acknowledge a stimulating environment at the Centre for Science and Technology Studies of the Leiden University where part of this paper was written.}


%
%

\appendix

\section{\label{app:nets}Networks and synthetic graphs}

In this appendix, we provide the details of the empirical networks and synthetic graphs analysed in the paper. The former include three examples of social collaboration networks, protein interactions networks, autonomous systems graphs and food webs. In all cases, the networks have been represented with simple undirected graphs and reduced to their largest connected components.

The collaboration networks represent jazz musicians that played in a band between 1912 and 1940 obtained from the Red Hot Jazz archive~\cite{GD03}, coauthorships between network scientists parsed from the bibliographies of two review papers in 2006~\cite{New06a} and coauthorships between Slovenian computer scientists up to 2010 extracted from the SICRIS database~\cite{BSB12}. The protein interactions networks represent recorded protein-protein interactions of the parasite \plasm, yeast \scere and nematode \celeg. The networks were collected from the BioGRID repository in 2016~\cite{SBRBBT06}. The autonomous systems graphs are the Internet maps at the level of autonomous systems~\cite{LKF07} on the first day of 1998, 1999 and 2000 reconstructed from the University of Oregon Route Views project~\cite{SNAP}. The food webs represent observed predator-prey relationships between the species of \littlerock~\cite{Mar91} and Florida Bay in wet and dry season~\cite{Pajek}.

\Tblref{nets} shows the standard statistics of the networks including the number of nodes $n$, the average node degree $\avg{k}=2m/n$, where $m$ is the number of edges, the average node clustering coefficient $\avg{C}$~\cite{WS98} with the clustering coefficient of node $i$ defined as $C_i=\frac{2t_i}{k_i(k_i-1)}$, where $t_i$ is the number of triangles including node $i$ and $k_i$ is its degree, the average distance between the nodes $\avg{\ell}=\frac{2}{n(n-1)}\sum_{i<j}d_{ij}$, where $d_{ij}$ is the number of edges in the shortest paths between nodes $i$ and $j$, and convexity $X$ defined in~\eqref{X} in~\secref{convex}.

\begin{table}[t]
	\caption{\label{tbl:nets}Statistics of empirical networks and synthetic graphs analysed in the paper. These show the number of nodes $n$, the average node degree $\avg{k}$ and clustering coefficient $\avg{C}$, the average distance between the nodes $\avg{\ell}$ and convexity $X$. The values are averages over $25$ independent~realisations.}
	\begin{tabular}{llrrrcc}
		Class & Network & \mc{$n$} & \mc{$\avg{k}$} & \mc{$\avg{\ell}$} & $\avg{C}$ & $X$ \\\hline
		\multirow{3}{*}{Collaboration} & \jazz & $198$ & $27.70$ & $2.24$ & $0.62$ & $0.12$ \\
		& \netsci & $379$ & $4.82$ & $6.04$ & $0.74$ & $0.85$ \\
		& \cmpsci & $239$ & $4.75$ & $4.58$ & $0.48$ & $0.64$ \\
		\multirow{3}{*}{\shortstack[l]{Protein\\ interactions}} & \plasm & $1158$ & $4.15$ & $4.24$ & $0.02$ & $0.43$ \\
		& \scere & $1458$ & $2.67$ & $6.81$ & $0.07$ & $0.68$ \\
		& \celeg & $3747$ & $4.14$ & $4.32$ & $0.06$ & $0.56$ \\	
		\multirow{3}{*}{\shortstack[l]{Autonomous\\ systems}} & \oreg{1998} & $3213$ & $3.50$ & $3.77$ & $0.18$ & $0.66$ \\
		& \oreg{1999} & $531$ & $4.58$ & $3.39$ & $0.18$ & $0.49$ \\
		& \oreg{2000} & $3570$ & $3.94$ & $3.80$ & $0.20$ & $0.59$ \\
		\multirow{3}{*}{Food webs} & \littlerock & $183$ & $26.60$ & $2.15$ & $0.32$ & $0.02$ \\
		& \baywet & $128$ & $32.42$ & $1.78$ & $0.33$ & $0.03$ \\
		& \baydry & $128$ & $32.91$ & $1.77$ & $0.33$ & $0.03$ \\\hline
		\multirow{9}{*}{\shortstack[l]{Synthetic\\ graphs}} & \multirow{3}{*}{Random graph} & $2500$ & $10.00$ & $3.65$ & $0.00$ & $0.00$ \\
		& & $1000$ & $10.00$ & $3.26$ & $0.01$ & $0.01$ \\
		& & $225$ & $10.00$ & $2.60$ & $0.05$ & $0.03$ \\
		& Triangular lattice & $225$ & $5.48$ & $8.07$ & $0.43$ & $0.23$ \\
		& Rectangular lattice & $225$ & $3.73$ & $10.00$ & $0.00$ & $0.13$ \\
		& Core-periphery graph & $3747$ & $4.48$ & $5.22$ & $0.00$ & $0.39$ \\
		& \multirow{3}{*}{Convex graph} & $2500$ & $5.97$ & $10.23$ & $0.90$ & $1.00$ \\
		& & $1000$ & $5.97$ & $8.40$ & $0.90$ & $1.00$ \\
		& & $225$ & $6.01$ & $5.44$ & $0.90$ & $1.00$
	\end{tabular} 
\end{table}

\Tblref{nets} also shows the statistics of the analysed synthetic graphs. First, these include triangular and rectangular lattices with the side of $15$ nodes and the Erd\H{o}s-R\'{e}nyi random graphs~\cite{ER59} with $n=225$, $1000$ or $2500$ nodes and the average degree $\avg{k}=10$ to ensure the graphs are almost fully connected. We generate simple graphs by forbidding any parallel edges or self-edges.

Next, core-periphery graphs are based on the core-periphery structure of the \celeg protein interactions network. Let $n=3747$ be the number of nodes in the network and $cn$ the number of nodes in its c-core as defined in~\secref{skeleton} with $c=0.43$. We generate random graphs with two sets of $cn$ and $(1-c)n$ nodes, and the probability of an edge between the nodes equal to the density within or between the network c-core and periphery. Because this leaves a non-trivial fraction of nodes isolated, any isolated node is reattached to the remaining nodes by a random edge. The average degree in the graphs is thus larger than in the corresponding network, $\avg{k}=4.48>4.14$. Nevertheless, the reattachment process actually creates pendant nodes that can only increase convexity in the generated graphs as in~\eqref{Xn1}, which is still lower than in the network, $X=0.39<0.56$.

Finally, convex graphs with $n\approx 225$, $1000$ or $2500$ nodes are constructed as follows. Since these are collections of cliques connected in a tree, we first generate a random tree on $tn$ nodes, where $t\in[2/n,1]$ is a parameter different than in the main text. Starting with a single node, we add $tn-1$ nodes one at a time, while each node is connected to a tree by a random edge. Then, every edge of the constructed tree is expanded to a clique on $k$ nodes by adding $k-2$ new nodes, where $k$ is sampled independently for each edge from the interval $[2,\frac{2(n-1)}{tn-1}]$. We set $t=0.25$ to match the characteristic size of cliques in empirical networks~\cite{SIL14}. Note, however, that this construction process cannot generate all possible convex graphs. The graphs also lack scale-free degree distribution~\cite{BA99}, although this could be imposed by an appropriate preferential attachment mechanism~\cite{HAMND11}.

\section{\label{app:random}Convexity under full randomisation}

In \secref{random}, we study the robustness of convexity in different empirical networks and synthetic graphs under degree-preserving randomisation by edge rewiring~\cite{MS02}. In this appendix, we consider also full randomisation, where endpoints of randomly selected edges are rewired to random other nodes. We ensure simple networks and graphs by forbidding any parallel edges or self-edges during the rewiring. At the limit, this process generates the Erd\H{o}s-R\'{e}nyi random graphs~\cite{ER59} that are non-convex.

\begin{figure}[h]
	\centering\includegraphics[width=0.6\textwidth]{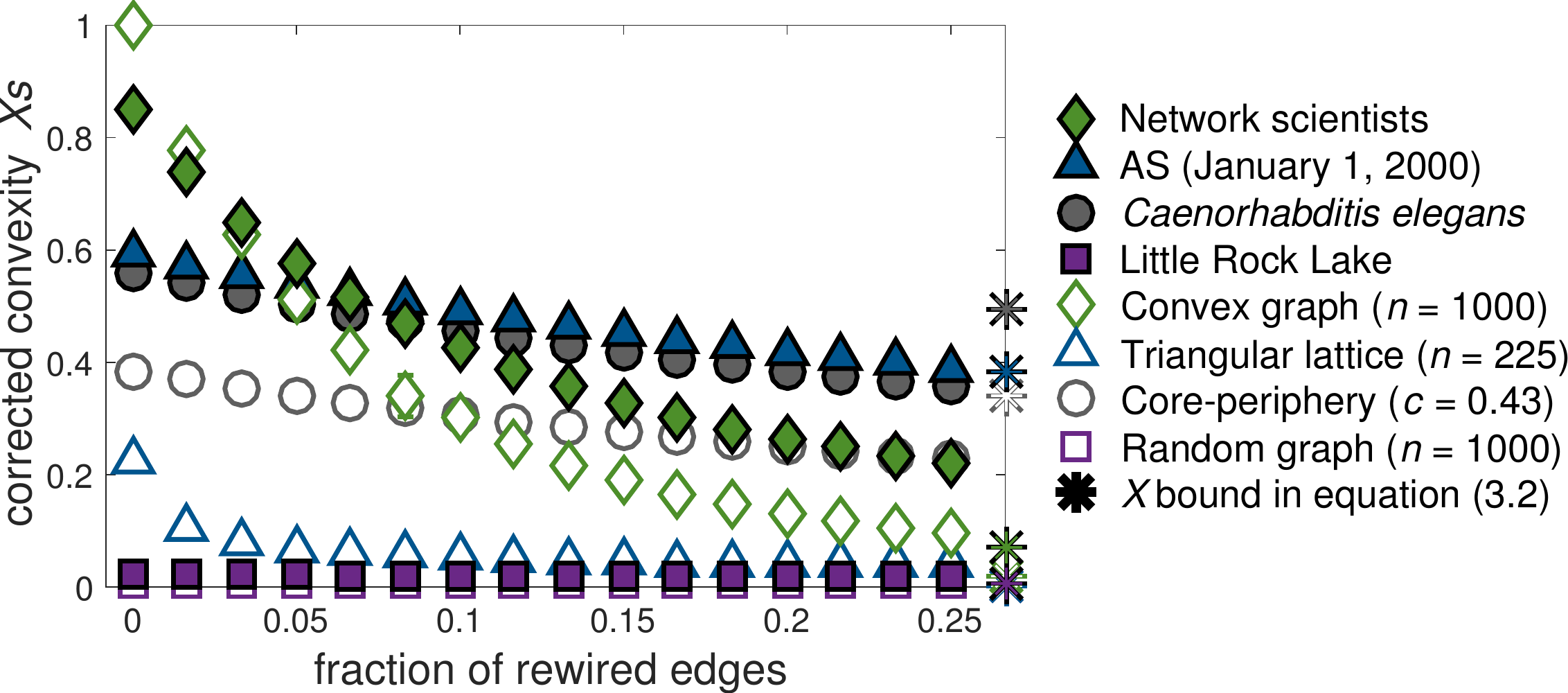}
	\caption{\label{fig:random}Evolution of convexity in empirical networks and synthetic graphs under full randomisation by edge rewiring, while other details are the same as in~\figref{rewiring}.} 
\end{figure}

\Figref{random} shows the evolution of corrected convexity $Xs$ in~\eqref{Xs} under full randomisation. The results are conceptually the same as for degree-preserving randomisation in~\figref{rewiring}, with the only difference that convexity $X$ is now no longer bound by~\eqref{Xn1}.

\section{\label{app:nodes}Node distributions of convex skeletons}

\begin{figure}[h]
	\centering\includegraphics[width=0.35\textwidth]{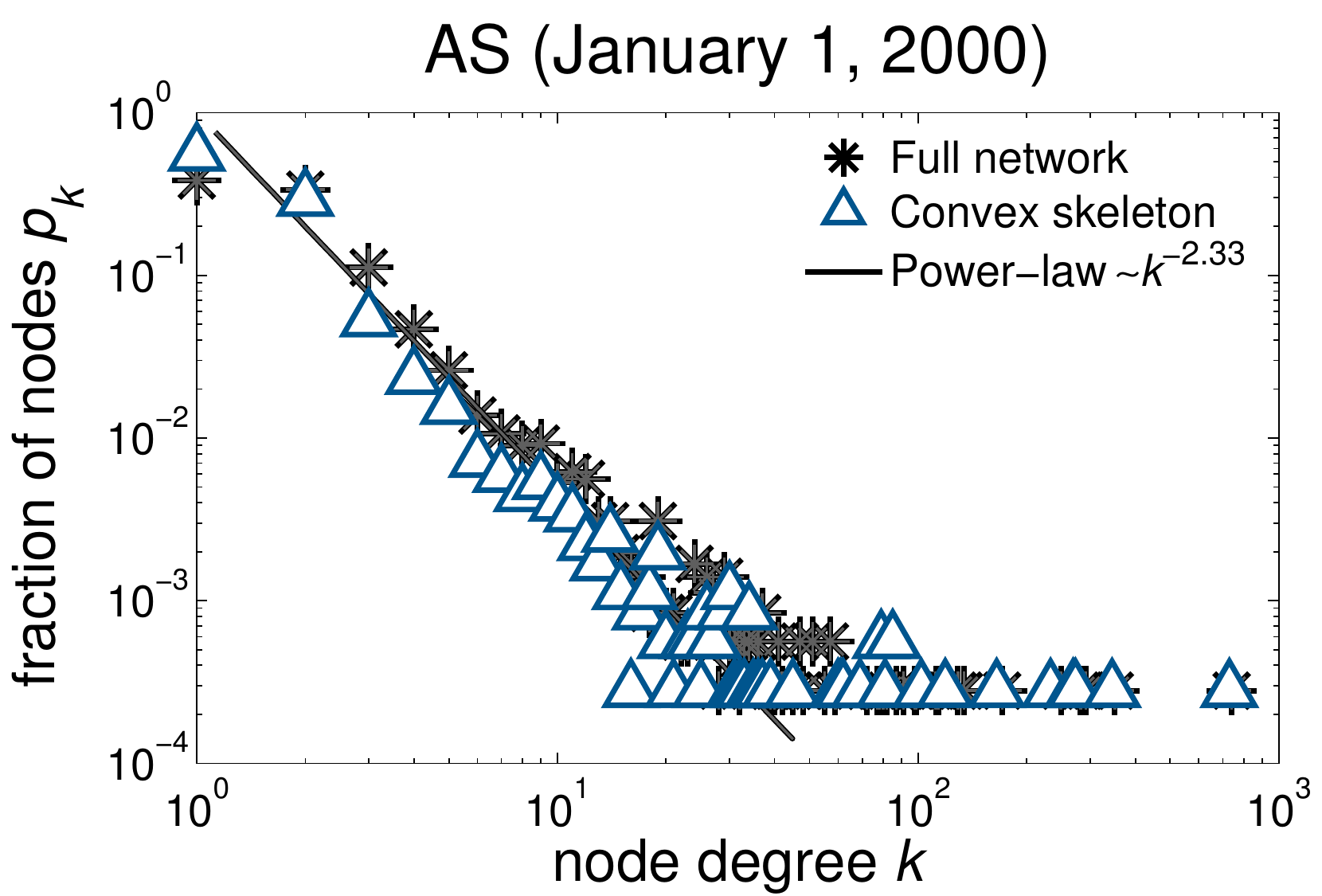}\hskip0.025\textwidth%
	\includegraphics[width=0.35\textwidth]{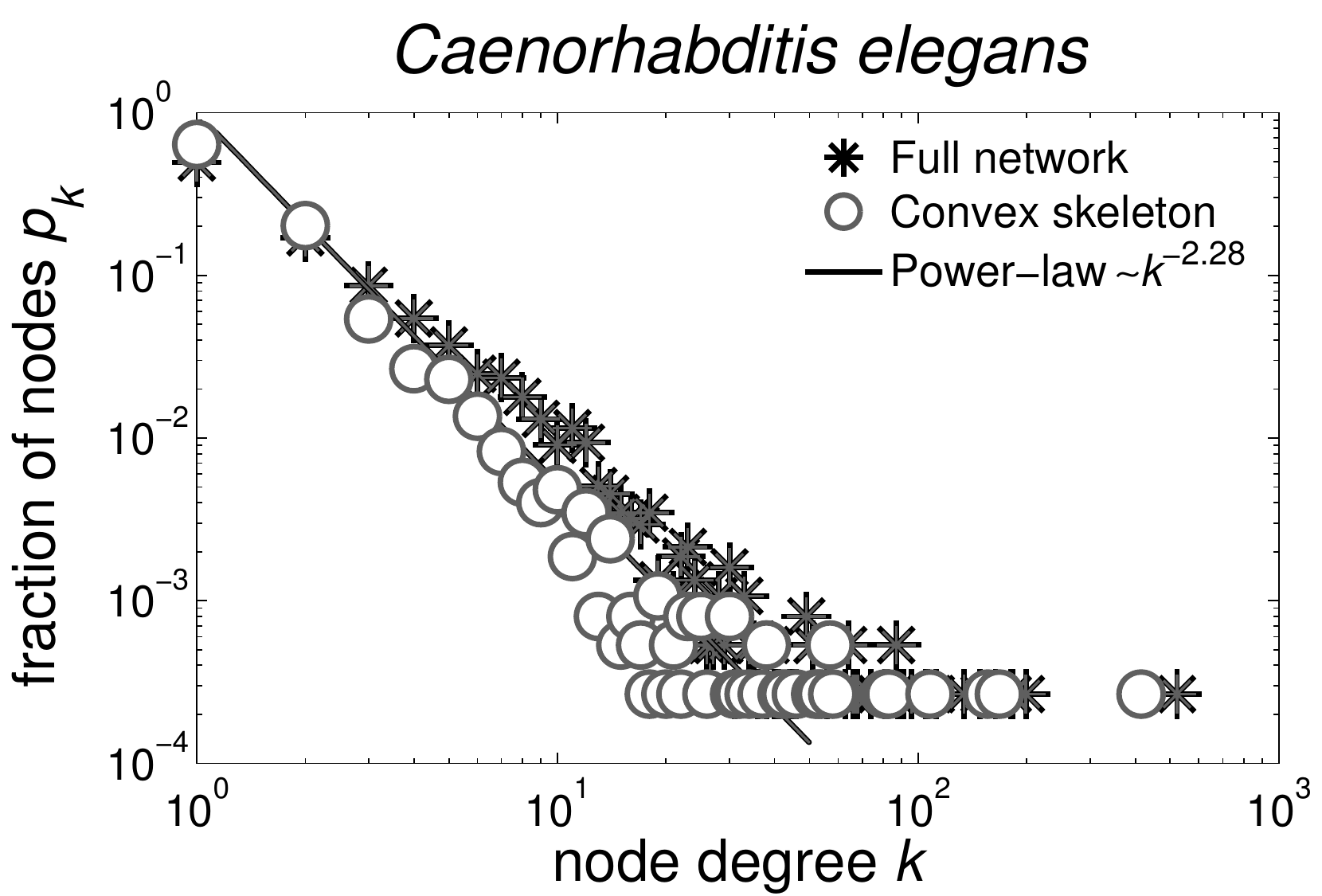}\\\vskip0.0125\textwidth%
	\includegraphics[width=0.35\textwidth]{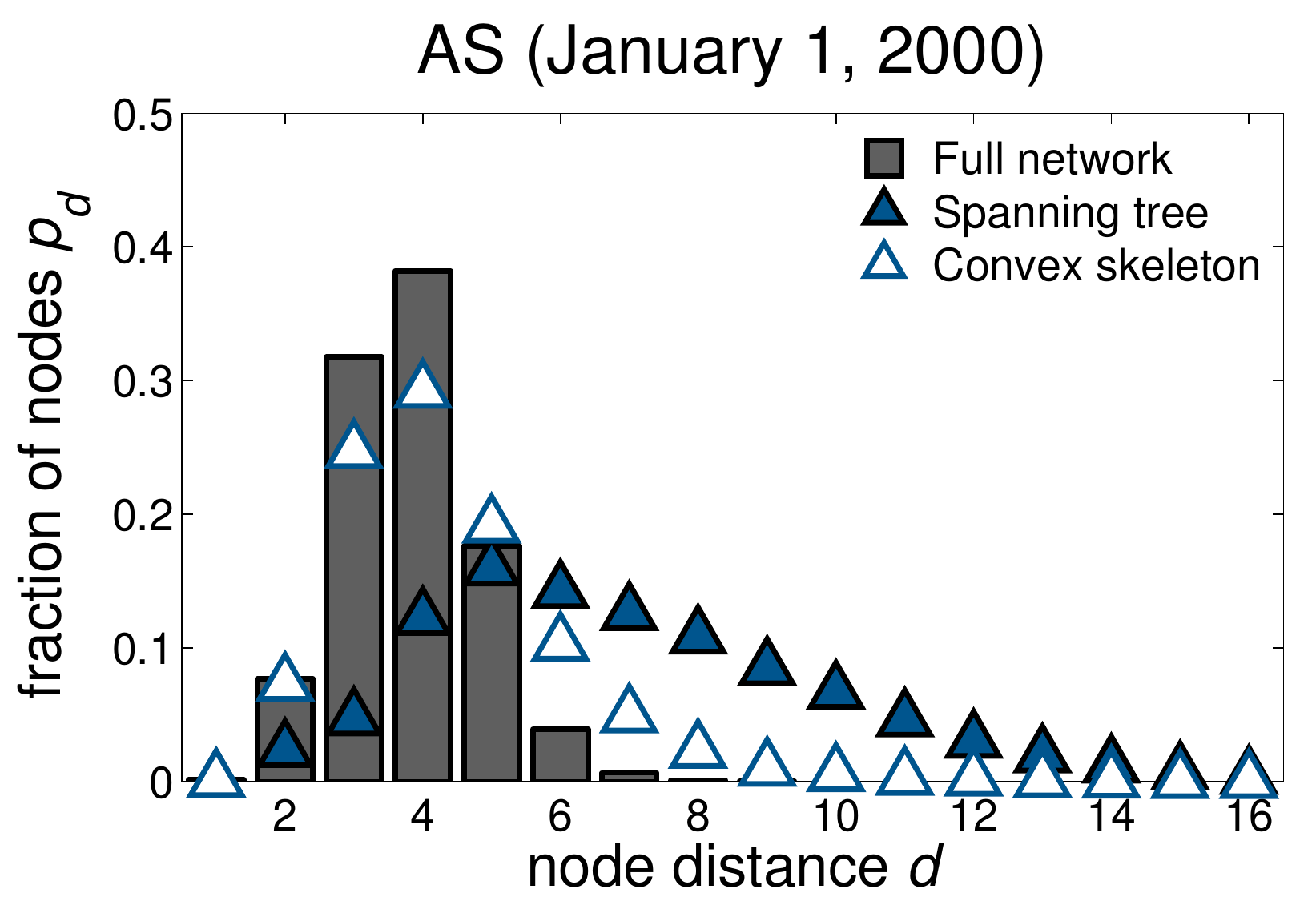}\hskip0.025\textwidth%
	\includegraphics[width=0.35\textwidth]{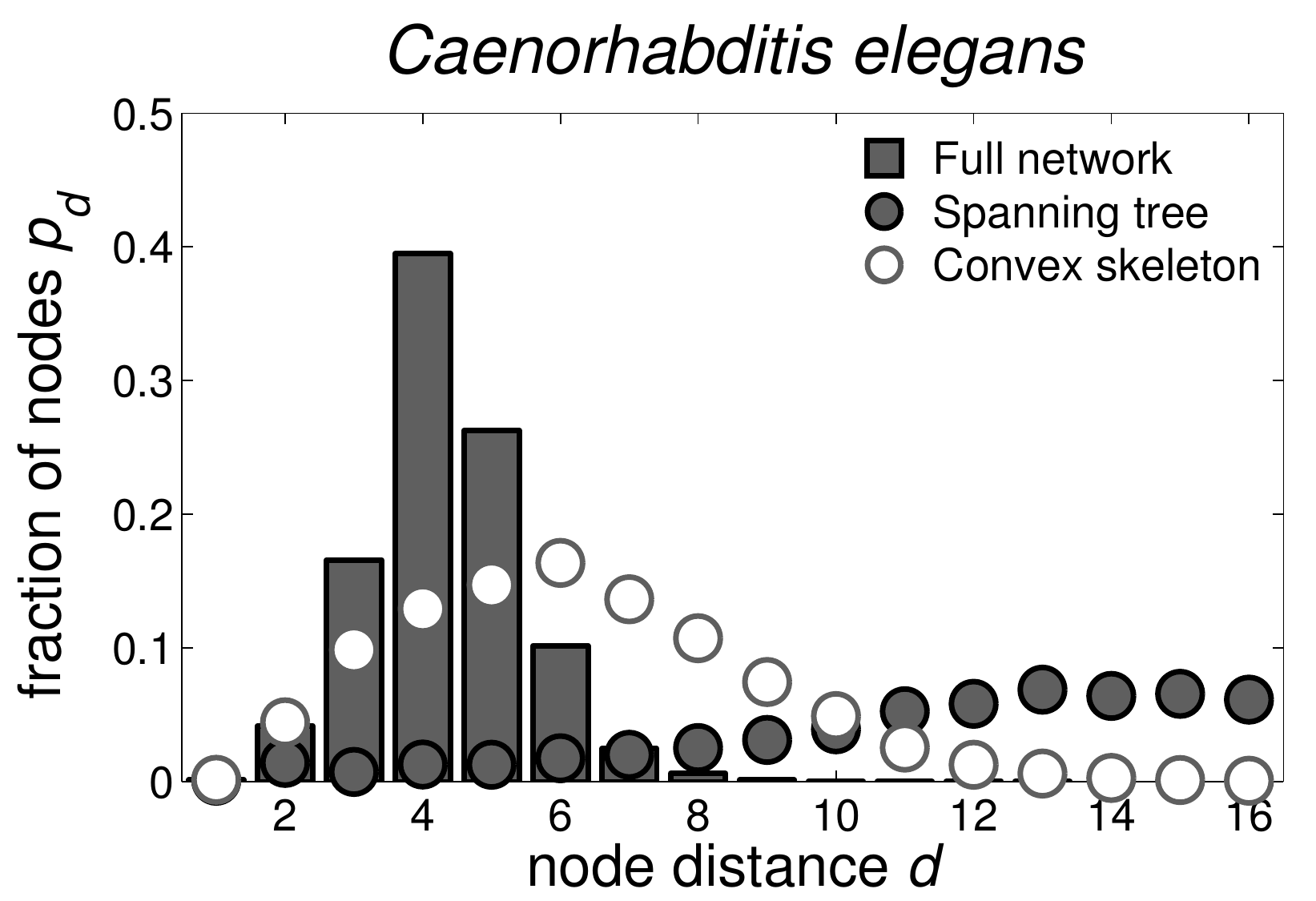}%
	\caption{\label{fig:nodes}Node distributions of core-periphery networks and particular realisations of spanning trees and convex skeletons, while other details are the same as in~\figref{skeleton}.} 
\end{figure}

In~\secref{skeleton}, we analyse different node distributions of convex skeletons and spanning trees extracted from smaller empirical networks. In this appendix, we consider also two larger networks, namely the autonomous systems graph from 2000 and the \celeg protein interactions network. The top row of~\figref{nodes} shows the node degree distributions $p_k$ of full networks (asterisks) and the extracted convex skeletons (\triane and \ellpe). These are largely consistent and seem to follow a power-law $k^{-\gamma}$ with $\gamma\approx 2.3$~\cite{CSN09}. Furthermore, the convex skeletons preserve the distributions of distances between the nodes $p_d$ shown in the bottom row of~\figref{nodes}. On the other hand, the distances between the nodes increase significantly in the spanning trees (\trians and \ellps).

%
%


\end{document}